\catcode`\@=11
\newif\if@fewtab\@fewtabtrue

{\count255=\time\divide\count255 by 60
\xdef\hourmin{\number\count255}
\multiply\count255 by-60\advance\count255 by\time
\xdef\hourmin{\hourmin:\ifnum\count255<10 0\fi\the\count255}}
\def\ps@draft{\let\@mkboth\@gobbletwo
    \def\@oddhead{}
    \def\@oddfoot
       {\hbox to 7 cm{$\scriptstyle Draft\ version:\ \draftdate$
       \hfil}\hskip -7cm\hfil\rm\thepage \hfil}
    \def\@evenhead{}\let\@evenfoot\@oddfoot}

\def\ceqno{\global\@fewtabfalse
    \ifcase\@eqcnt \def\@tempa{& & &}\or \def\@tempa{& &}
      \or \def\@tempa{&}
      \or\def\@tempa{}\fi\@tempa
{\rm(\theequation)}}

\def\aeqno#1{\global\@fewtabfalse
    \ifcase\@eqcnt \def\@tempa{& & &}\or \def\@tempa{& &}
      \or \def\@tempa{&}
      \or\def\@tempa{}\fi\@tempa
{\rm(\theequation,#1)}}

\def\label#1{\ifnum\draftcontrol=1
 \global\def\draftnote{$\scriptstyle #1$}\fi
 \@bsphack\if@filesw {\let\thepage\relax
   \def\protect{\noexpand\noexpand\noexpand}%
\xdef\@gtempa{\write\@auxout{\string
      \newlabel{#1}{{\@currentlabel}{\thepage}}}}}\@gtempa
   \if@nobreak \ifvmode\nobreak\fi\fi\fi
  \@esphack}

\def\alabel#1#2{\label{#1}\global\@fewtabfalse
    \ifcase\@eqcnt \def\@tempa{& & &}\or \def\@tempa{& &}
      \or \def\@tempa{&}
      \or\def\@tempa{}\fi\@tempa
{\hbox to 3cm{\phantom{\rm(\theequation,#2)}
\draftnote \hfil}\hskip -3cm {\rm(\theequation,#2)}}}

\def\clabel#1{\label{#1}\global\@fewtabfalse
    \ifcase\@eqcnt \def\@tempa{& & &}\or \def\@tempa{& &}
      \or \def\@tempa{&}
      \or\def\@tempa{}\fi\@tempa
{\hbox to 3cm{\phantom{\rm(\theequation)}
\draftnote \hfil}\hskip -3cm{\rm(\theequation)}}}

\def\eqnarray{\def\draftnote{{}}\global\@fewtabtrue
\stepcounter{equation}\let\@currentlabel=\theequation
\global\@eqnswtrue
\global\@eqcnt\z@\tabskip\@centering\let\\=\@eqncr
$$\halign to \displaywidth\bgroup\@eqnsel\hskip\@centering\@eqcnt\z@
  $\displaystyle\tabskip\z@{##}$&\global\@eqcnt\@ne
  \hskip 1\arraycolsep \hfil${##}$\hfil
  &\global\@eqcnt\tw@ \hskip 1\arraycolsep
$\displaystyle\tabskip\z@{##}$
\hfil  \tabskip\@centering&\global\@eqcnt\thr@@\llap{##}\tabskip\z@
\cr}

\def\endeqnarray{\@@eqncr\egroup
      \global\advance\c@equation\m@ne$$\global\@ignoretrue}

\def\@eqnnum{\hbox to 3cm{\phantom{\rm(\theequation)} \draftnote
                         \hfil}\hskip -3cm {\rm(\theequation)}}

\def\@@eqncr{\let\@tempa\relax
    \ifcase\@eqcnt \def\@tempa{& & &}\or \def\@tempa{& &}
      \or \def\@tempa{&}
      \or\def\@tempa{}
\fi\@tempa
\if@eqnsw
\if@fewtab\@eqnnum\fi
\stepcounter{equation}\fi\global
\@eqnswtrue\global\@eqcnt\z@\global\@fewtabtrue\cr}

\def\draftcite#1{\ifnum\draftcontrol=1#1\else{}\fi}

\def\@lbibitem[#1]#2{\item{}\hskip -3cm \hbox to 2cm
{\hfil$\scriptstyle\draftcite{#2}$}\hskip
1cm[\@biblabel{#1}]\if@filesw
     {\def\protect##1{\string ##1\space}\immediate
      \write\@auxout{\string\bibcite{#2}{#1}}}\fi\ignorespaces}

\def\@bibitem#1{\item\hskip -3cm \hbox to 2cm
{\hfil $\scriptstyle\draftcite{#1}$}\hskip 1cm
\if@filesw \immediate\write\@auxout
       {\string\bibcite{#1}{\the\value{\@listctr}}}\fi\ignorespaces}

     \def\nsection#1{\section{#1}\setcounter{equation}{0}}
     
     \def\nappendix#1{\vskip 1cm\no{\bf Appendix
         #1}\def\thesection{#1} \setcounter{equation}{0}}

\newfam\dlfam  % \dl is double line
\newfam\glfam  % \gl is gothic letters
\def\draftdate{\number\month/\number\day/\number\year\ \ \ \hourmin }
 \global\def\draftcontrol{0}
\catcode`\@=12 \def\tilde{\widetilde} \def\hat{\widehat}
\documentstyle[12pt]{article}
\input{psfig}

\def\theequation{{\thesection.\arabic{equation}}}

\newcommand{\be}{\begin{eqnarray}} \newcommand{\en}{\end{eqnarray}\vs
  0.5 cm}  \newcommand{\no}{\noindent}
\newcommand{\vs}{\vskip} 
 
 \newcommand{\un}{\underline}
\newcommand{\NR}{{{\bf R}}}%letra doble raya en modo matematico
\newcommand{\Nx}{{{\bf x}}} 
\newcommand{\Nv}{{{\bf v}}}

 \newcommand{\Nr}{{{\bf r}}}
\newcommand{\qq}{\begin{eqnarray}} 
  \newcommand{\da}{\partial} \newcommand{\ee}{{\rm e}}
   \newcommand{\qqq}{\end{eqnarray}}

 \newcommand{\CC}{{\cal C}}

 \newcommand{\CO}{{\cal O}}
\newcommand{\CP}{{\cal P}} 
 \newcommand{\CS}{{\cal S}}

\newcommand{\CZ}{{\cal Z}} 
\newcommand{\m}{\hspace{0.025cm}} 
 
\newcommand{\hf}{{_1\over^2}}

\begin{document}
\title{\bf{Phase transition in the passive scalar advection}}

\author{Krzysztof Gaw\c{e}dzki \\C.N.R.S., I.H.E.S.,
  91440 Bures-sur-Yvette, France\\
  \\
  Massimo Vergassola \\C.N.R.S., Observatoire de la C\^ote d'Azur, B.P. 4229,\\
  06304 Nice, France}
\date{ } 
\maketitle
\vskip 0.2cm
\begin{abstract}
\vskip 0.2cm
\noindent The paper studies the behavior of the trajectories 
of fluid particles in a compressible generalization of the Kraichnan 
ensemble of turbulent velocities. We show that, depending on the 
degree of compressibility, the trajectories either explosively
separate or implosively collapse. The two behaviors are shown
to result in drastically different statistical properties of scalar 
quantities passively advected by the flow. At weak compressibility,
the explosive separation of trajectories induces a familiar direct 
cascade of the energy of a scalar tracer with a short-distance 
intermittency and dissipative anomaly. At strong compressibility, 
the implosive collapse of trajectories leads to an inverse cascade 
of the tracer energy with suppressed intermittency and with the energy 
evacuated by large scale friction. A scalar density whose advection
preserves mass exhibits in the two regimes opposite cascades
of the total mass squared. We expect that the explosive separation
and collapse of Lagrangian trajectories occur also in more
realistic high Reynolds number velocity ensembles and that the
two phenomena play a crucial role in fully developed 
turbulence.
\vskip 0.3cm

\hfill
\end{abstract}

%%%%%%%%%%%%%%%%%%%%%%%%%%%%%%%%%%%%%%%%%%%%%%%%%%%%%%%%%%%%%%%%
%%% for draft versions, suppress in definitive version
%\draft
%%
%%%%%%%%%%%%%%%%%%%%%%%%%%%%%%%%%%%%%%%%%%%%%%%%%%%%%%%%%%%%%%%%

\vs 1cm

\nsection{Introduction}

One of the main characteristic features of the high Reynolds number
turbulent flows is a cascade-like transfer of the energy injected by
an external source. In three dimensional flows, the injected energy is
transferred to shorter and shorter scales and is eventually dissipated
by the viscous friction.  This direct cascade is in the first
approximation described by the Kolmogorov 1941 scaling theory
\cite{K41} but the observed departures from scaling (intermittency)
remain to be explained from the first principles.  As discovered by
R.H.~Kraichnan in \cite{Kr67}, in two dimensions, the injected energy
is transferred to longer and longer distances in an inverse cascade
whereas this is the enstrophy that is transferred to shorter and
shorter scales. Experiments \cite{Tab} and numerical simulations
\cite{SY,Betal} suggest the absence of intermittency in the inverse
2-dimensional cascade.  In the present paper, we shall put forward
arguments indicating that the occurrence and the properties of direct
and inverse cascades of conserved quantities in hydrodynamical flows
are related to different typical behaviors of fluid particle
trajectories.  \vskip 0.3cm

Let us start by drawing some simple analogies between fluid dynamics
and the theory of dynamical systems which studies solutions of the
ordinary differential equations \qq {{dx}\over{dt}}\,=\, X(x)\m.
\label{ode}
\qqq Let $x_{_{s,y}}(t)$ denote the solution of Eq.\,\,(\ref{ode})
passing at time $s$ by point $y$.  In dynamical systems, where the
attention is concentrated on regular functions $X$, one encounters
different types of behavior of solutions\footnote{the following is not
  a statement about the genericity of the listed behaviors} \vskip
0.2cm

1). \ {\bf integrable motions} (more common in Hamiltonian systems),
where the nearby trajectories stay close together forever: \qq
\label{one}
\vert x_{_{s,y_1}}(t)-\m x_{_{s,y_2}}(t)\vert\ \sim\ \CO(\vert
y_1-y_2\vert)\m,\ \qquad \qqq \vskip 0.1cm

2). \ {\bf chaotic motions} where the distance between the nearby
trajectories grows exponentially, signaling a sensitive dependence on
the initial conditions: \qq
\label{two}
\vert x_{_{s,y_1}}(t)-\m x_{_{s,y_2}}(t)\vert\ \sim\ 
\CO(\ee^{\lambda\vert t-s\vert}\vert y_1-y_2\vert)\m,\, \qqq with the
Lyapunov exponent $\lambda>0$\m, \vskip 0.2cm

3). \ last but, by no means, least, {\bf dissipative motions} where
\qq
\label{three}
\vert x_{_{s,y_1}}(t)-\m x_{_{s,y_2}}(t)\vert\ \sim\ 
\CO(\ee^{\lambda\vert t-s\vert}\vert y_1-y_2\vert)\m, \qqq with
$\lambda<0$.  \vskip 0.2cm

\noindent Various of these types of motions may appear in 
the same systems.  \vskip 0.3cm

Analogies between dynamical systems and hydrodynamical evolution 
equations, for example the Navier Stokes ones, are often drawn 
by viewing the Eulerian evolution of velocities as a dynamical 
system in infinite dimensions, see \cite{RuTak}. One
has, however, a more direct (although not unrelated) analogy between
Eq.\,\,(\ref{ode}) and the ordinary differential equation \qq
{{d\Nx}\over{dt}}\ =\ \Nv(t,\Nx)
\label{odeh}
\qqq for the Lagrangian trajectories of fluid particles in a given
velocity field $\Nv(t,\Nx)$. As before, we shall denote by
$\Nx_{_{s,\Nr}}(t)$ the solution passing by $\Nr$ at time $s$.
Clearly, the system (\ref{odeh}) is time-dependent and the velocity
field is itself a dynamical variable. Nevertheless, one may ask
questions about the behavior of solutions of Eq.\,\,(\ref{odeh}) for
"typical" velocities.  On the phenomenological level, such behavior
seems to be rather robust and to depend on few characteristics of the
velocity fields. One of them is the Reynolds number
$Re={L\m\vert\Delta_{_L} \Nv\vert\over\nu}$, where
$\vert\Delta_{_L}\Nv\vert$ is the (typical) velocity difference over
the distance $L$ of the order of the size of the system and $\nu$ is
the kinematic viscosity. Another important characteristic of velocity
fields is the degree of compressibility measured, for example, by the
ratio of mean values of $(\sum\limits_\alpha\nabla_\alpha
v^\alpha)^2\equiv (\nabla\cdot\Nv)^2$ and $\sum\limits_{\alpha,\beta}
(\nabla_\alpha v^\beta)^2\equiv(\nabla\Nv)^2$.  \vskip 0.2cm

Reynolds numbers ranging up to $\CO(10^2)$ are the realm of
laminar flows and the onset of turbulence. Velocity fields in
(\ref{odeh}) are thus regular in space and the behaviors
(1) to (3) are observed for Lagrangian trajectories.
They seem to have limited bearing on the character of the Eulerian 
evolution of velocities, see Chapter 8 of \cite{BJPV}.
This is a natural domain of applications of the theory of dynamical
systems to both Eulerian and Lagrangian evolutions. 
When the Reynolds number is increased, however, fully
developed turbulent flows are produced in which the behavior of
trajectory separations becomes more dramatic. For incompressible
flows, for example, we claim (see also
\cite{slowm,fmv,nice}) that the regime of fully
developed turbulence is characterized by the
\vskip 0.2cm 

2'). {\bf explosive separation of trajectories}: 
\qq
\qquad\qquad\vert\Nx_{_{s,\Nr_1}}(t)-\m\Nx_{_{s,\Nr_2}}(t)\vert \quad\ 
{\rm becomes\ }\CO(1)\ {\rm in\ finite\ time}\m.
\label{2prim}
\qqq More precisely, the time of separation of trajectories to an
$\CO(1)$ distance is bounded when $y_2$ approaches $y_1$, provided
that the initial separation $\vert y_1-y_2\vert$ stays in the inertial
range where the viscous effects may be neglected.  \vskip 0.3cm

Since the inertial range extends down to zero distance when
$Re\to\infty$, the fast separation of trajectories has a drastic
consequence in this limit: the very concept of individual Lagrangian
trajectories breaks down.  Indeed, at $Re=\infty$, infinitesimally
close trajectories take finite time to separate\footnote{in contrast
to their behavior in the chaotic regime, see Sect.\,\,2.2 below}
and, as a result, there are many trajectories (in fact, a continuum)
satisfying a given initial condition. It should be still possible,
however, to give a statistical description of such ensembles of
trajectories in a fixed velocity field. Unlike for intermediate 
Reynolds numbers, there seems to be a strong relation between 
the behavior of the Lagrangian trajectories and the basic hydrodynamic 
properties of developed turbulent flows: we expect the appearance 
of non-unique trajectories for $Re\to\infty$ to be responsible 
for the dissipative anomaly, the direct energy cascade, 
the dissipation of higher conserved quantities and the pertinence 
of weak solutions of hydrodynamical equations at $Re=\infty$.   
\vskip 0.3cm

The breakdown of the Newton-Leibniz paradigm based on the uniqueness
of solutions of the initial value problem for ordinary differential
equations is made mathematically possible by the loss of small scale
smoothness of turbulent velocities when $Re\to\infty$. At $Re=\infty$,
the typical velocities are expected to be only H\"{o}lder continuous
in the space variables: \qq \vert\Nv(t,\Nx)-\Nv(t,\Nx')\vert^2\ \sim\ 
\vert \Nx-\Nx'\vert^{\xi}\m,
\label{hoel}
\qqq with the H\"{o}lder exponent ${\xi\over2}$ close to the Kolmogorov 
value ${1\over3}$ \cite{K41}. The uniqueness of solutions of the initial 
value problem for Eq.\,\,(\ref{odeh}) requires, on the other hand, 
the Lipschitz continuity of $\Nv(t,\Nx)$ in $\Nx$, i.e. the behavior
(\ref{hoel}) with $\xi=2$. It should be stressed that for large but
finite $Re$, the chaotic behavior (2) of trajectories may still
persist for short separations of the order of the dissipative scale
(where the viscosity makes the velocities smooth) and the behavior
(2') is observed only on distances longer than that. However, it is
the latter which seems responsible for much of the observed physics of
fully developed turbulence and, thus, setting $Re=\infty$ seems to
be the right idealization in this regime.  \vskip 0.3cm

For general velocity fields, one should expect that the poor spatial
regularity of velocities $\Nv$ might lead to two opposite effects.  On
one hand, the trajectories may branch at every time and coinciding
particles would split in a finite time as in (2'). Solving discretized
versions of Eq.\,\,(\ref{odeh}) randomly picks a branch of the
solution and generates some sort of a random walk whose average
reproduces the trajectory statistics. This is the effect previously
remarked \cite{slowm,fmv,nice} in the studies of the
incompressible Kraichnan model \cite{Kr94}. It should be dominant for
incompressible or weakly compressible flows. On the other hand, the
trajectories may tend to be trapped together. The most direct way to
highlight this phenomenon is to consider strongly compressible
velocity fields which are well known for depleting transport (see
\cite{VA}). An instance is provided by the one-dimensional equation
$\,{dx\over dt}\m=\m\beta(x)\m,\,$ for $\beta(x)$ a Brownian motion in
$x$, whose solutions are trapped in finite time at the zeros of
$\beta$ on the right (left) ends of the intervals where $\beta>0$
($\beta<0$) \cite{sinai2}.  
\vskip 0.3cm

What may then become typical in strongly compressible velocities is, 
instead of the explosion (2'), the
\vskip 0.2cm

3'). {\bf implosive collapse of trajectories}: 
\qq
\qquad\qquad\vert\Nx_{_{s,\Nr_1}}(t)-\m\Nx_{_{s,\Nr_2}}(t) \vert\quad\ 
{\rm becomes \ equal\ to\ zero\ in\ finite\ time}\m, \qqq with the
time of collapse depending on the initial positions. This type of
behavior should lead to a domination at infinite $Re$ and strong
compressibility of shock-wave-type solutions of hydrodynamical
equations, as in the 1-dimensional Burgers problem \cite{burg}. Again,
strictly speaking, we should expect behavior (3') only for $Re=\infty$
whereas for finite $Re$, on distances smaller than the dissipative
scale, the approach of typical trajectories should become exponential
with a negative Lyapunov exponent. In simple Gaussian ensembles 
of smooth compressible velocities the latter behavior and its
consequences for the direction of the cascade of a conserved quantity
have been discovered and extensively discussed in \cite{CKV1} and
\cite{CKV2}.  \vskip 0.3cm

It is the main purpose of the present paper to provide some support
for the above, largely conjectural, statements about typical behaviors
of fluid-particle trajectories at high Reynolds numbers and about 
the impact of these behaviors on physics of the fully turbulent 
hydrodynamical flows. We study only simple
synthetic random ensembles of velocities showing H\"{o}lder continuity
in spatial variables. Although this is certainly insufficient to make
firm general statements, it shows, however, that the behaviors (2') 
and (3') are indeed possible and strongly affect hydrodynamical properties. 
In realistic flows, both behaviors might coexist. For the ensemble of
flows considered here, they occur alternatively, leading to
two different phases at large (infinite) Reynolds numbers, depending
on the degree of compressibility and the space dimension. The
occurrence of the collapse (3') is reflected in the suppression
of the short-scale dissipation and the inverse cascade of certain 
conserved quantities. The absence of dissipative anomaly permits 
an analytical understanding of the dynamics and to show
that the inverse cascade is self-similar. This strengthens 
the conjectures on the general lack of intermittency for inverse 
cascades \cite{Tab,SY,Betal,Bisk}. A {\it caveat} comes however 
from the consideration of friction effects, indicating that the role 
of infrared cutoffs might be subtle and anomalies might reappear 
in terms of them.

\vskip 0.3cm

Synthetic ensembles of velocities are often used to study the problems
of advection in hydrodynamical flows. These problems become then
simpler to understand than the advection in the Navier-Stokes flows,
but might still render well some physics of the latter, especially in
the case of passive advection when the advected quantities do not
modify the flows in an important way.  The simplest of the problems
concern the advection of scalar quantities. There are two types of
such quantities that one may consider. The first one, which we shall
call a tracer and shall denote $\theta(t,\Nr)$, undergoes the
evolution governed by the equation \qq \da_t
\theta+\Nv\cdot\nabla\theta-\kappa\nabla^2\theta =f\m,
\label{ps}
\qqq where $\kappa$ denotes the diffusivity and $f(t,\Nr)$ describes
an external forcing (a source).  The second scalar quantity is of a
density type, e.g. the density of a pollutant, and we shall denote it
by $\rho(t,\Nr)$. Its evolution equation is \qq \da_t
\rho+\nabla\cdot(\Nv\m\rho)-\kappa\nabla^2\rho =f\m.
\label{psb}
\qqq For $f=0$ it has a form of the continuity equation so that,
without the source, the evolution of $\rho$ preserves the total mass
$\int\hspace{-0.08cm}\rho(t,\Nr)\m d\Nr$. Both equations coincide 
for incompressible velocities but are different if the velocities 
are compressible.
\vskip 0.3cm

A lot of attention has been attracted recently by a theoretical and
numerical study of a model of the passive scalar advection introduced
by R. H. Kraichnan in 1968 \cite{Kr68}.  The essence of the Kraichnan
model is that it considers a synthetic Gaussian ensemble of velocities
decorrelated in time.  The ensemble may be defined by specifying the
1-point and the 2-point functions of velocities. Following Kraichnan,
we assume that the mean velocity $\langle\m\Nv(t,\Nx)\m\rangle$
vanishes and that \qq \langle v^\alpha(t,\Nr)\,
v^\beta(t',\Nr')\rangle =\delta(t-t')\,[\m
d_0^{\alpha\beta}-d^{\alpha\beta}(\Nr-\Nr')\m]
\label{vc}
\qqq with constant $d_0^{\alpha\beta}$ and with $d^{\alpha\beta}(\Nr)$
proportional to $r^\xi$ at short distances.  The latter property
mimics the scaling behavior of the equal-time velocity correlators of
realistic turbulent flows in the $Re\to\infty$ limit.  It leads to the
behavior (\ref{hoel}) for the typical realizations of $\Nv$. The
parameter $\xi$ is taken between $0$ and $2$ so that the typical
velocities of the ensemble are not Lipschitz continuous.  
\vskip 0.3cm

We shall study the behavior of Lagrangian trajectories in the velocity
fields of a compressible version of the Kraichnan ensemble and the
effect of that behavior on the advection of the scalars. The time
decorrelation of the velocity ensemble is not a very physical
assumption. It makes, however, an analytic study of the model much
easier. We expect the temporal behavior of the velocities to have less
bearing on the behavior of Lagrangian trajectories than the
spatial one, but this might be the weakest point of our arguments.
Another related weak point is that the Kraichnan ensemble is
time-reversal invariant whereas realistic velocity ensembles are not,
so that the typical behaviors of the forward- and backward-in-time
solutions of Eq.\,\,(\ref{odeh}) for the Lagrangian trajectories may
be different. It should be also mentioned that in our conclusions
about the advection of scalars we let $Re\to\infty$ before sending the
diffusivity $\kappa$ to zero, i.e. we work at zero Schmidt or Prandtl
number $\nu\over\kappa$. The qualitative picture should not be
changed, however, if $Re$ becomes very large but $\nu\over\kappa$
stays bounded. On the other hand, the situation when
${\nu\over\kappa}\to\infty$ should be better described by the
$\xi\to2$ limit of the Kraichnan model where the velocities become
smooth.  \vskip 0.3cm

The paper is organized as follows. In Sect.\,\,2 we discuss the
statistics of the Lagrangian trajectories in the Kraichnan ensemble
and discover two different phases with typical behaviors (2') and
(3'), occurring, respectively, in the case of weak and strong
compressibility. In Sects.\,\,3 and 4 we discuss the advection of a
scalar tracer in both phases and show that it exhibits cascades of the
mean tracer energy.  In the weakly compressible phase, the cascade is
direct (i.e. towards short distances) and it is characterized by an
intermittent behavior of tracer correlations, signaled by their
anomalous scaling at short distances.  On the other hand, in the
strongly compressible phase, the tracer energy cascade inverts its
direction.  In the latter case, we compute exactly the probability
distribution functions of the tracer differences over long distances
and show that, although non-Gaussian, they have a scale-invariant
form. This indicates that the inverse cascade directed towards long
distances forgets the integral scale where the energy is injected.
Conversely, the scale where friction extracts energy from the system
is shown in Section~6 to lead to anomalous scaling of certain
observables. Finally, in Section 7, we discuss briefly the advection
of a scalar density.  Here, in the weakly compressible phase, we find
a cascade of the mean mass squared towards long distances and, on
short distances, the scaling of correlation functions in agreement
with the predictions of \cite{russ}. The strongly compressible
shock-wave phase, however, exhibits a drastically different behavior
with the inversion of the direction of the cascade of mean mass
squared towards short distances. In Conclusions we briefly summarize
our results. Some of the more technical material is
assembled in five Appendices.  
\vskip 0.3cm

\nsection{Lagrangian flow}

The assumptions of isotropy and of scaling behavior on all scales fix
the functions $d^{\alpha\beta}(x)$ in the velocity 2-point function
(\ref{vc}) of the Kraichnan ensemble up to two parameters: \qq
d^{\alpha\beta}(\Nr)=[A+(d+\xi-1)B]\,\delta^{\alpha\beta} \m
r^\xi\,+\,[A-B]\,\xi\m r^\alpha r^\beta\m r^{\xi-2}\m,
\label{d}
\qqq with $A=0$ corresponding to the incompressible case where
$\nabla\cdot\Nv=0$ and $B=0$ to the purely potential one with
$\Nv=\nabla\phi$.  Positivity of the covariance requires that
$A,B\geq 0$.  It will be convenient to relabel the constants $A$ and
$B$ by $\CS^2=A+(d-1)B$ and $\CC^2=A$. $\CS^2$ and $\CC^2$ are
proportional to, respectively, $\langle (\nabla\Nv)^2\rangle$ and
$\langle (\nabla\cdot\Nv)^2\rangle$ and they satisfy the inequalities
$\CS^2\geq\CC^2\geq0$. In one dimension, $\CS^2=\CC^2\geq0$. The ratio
\qq
\wp\equiv{\CC^2\over\CS^2},\quad 0\leq\wp\leq 1,
\label{iota}
\qqq
characterizes the
degree of compressibility.  \vskip 0.3cm

The source $f$ in the evolution equations (\ref{ps}) and (\ref{psb})
for the scalars will be also taken random Gaussian, independent of
velocity, with mean zero and 2-point function \qq \langle f(t,\Nr)\,
f(t',\Nr')\rangle =\delta(t-t')\, \chi({\vert\Nr-{\Nr}'\vert})\m,
\label{fc}
\qqq where $\chi$ decays on the injection scale $L$.  \vskip 0.3cm

In the absence of the forcing and diffusion terms in
Eq.\,\,(\ref{ps}), the tracer $\theta$ is carried by the flow:
$\theta(t,\Nr)\,=\,\theta(s,\Nx_{_{t,\Nr}}(s))$, and the density
$\rho$ is stretched as the Jacobian $J={\da(\Nx_{t,\Nr}(s))}/{\da(\Nr)}$ 
of the map $\Nr\mapsto\Nx_{t,\Nr}(s)$: 
$\rho(t,\Nr)\,=\,\rho(s,\Nx_{_{t,\Nr}}(s))\,J$. 
Here, $\Nx_{t,\Nr}(s)$ is the fluid
(Lagrangian) trajectory obeying $d\Nx_{_{t,\Nr}}/ds=v(s,\Nx_{_{t,\Nr}})$ 
and passing through the point $\Nr$ at time
$t$, i.e.  $\Nx_{_{t,\Nr}}(t)=\Nr$.  The flows of the scalars may be
rewritten as the relations \qq
\theta(t,\Nr)=\int\delta(\Nr'-\Nx_{_{t,{\bf r}}}(s))\,
\theta(s,\Nr')\,\m d\Nr'\m,\qquad
\rho(t,\Nr)=\int\delta(\Nr-\Nx_{_{s,{\bf r}'}}(t))\, \rho(s,\Nr')\,\m
d\Nr'\quad
\label{sc00}
\qqq which imply that the two flows are dual to each other:
$\,\int\theta(t,\Nr)\,\rho(t,\Nr)\,d\Nr$ does not change in time.
In the presence of forcing and diffusion, there are some slight
modifications.  First, the sources create the scalars along the
Lagrangian trajectories. Second, the diffusion superposes Brownian
motions upon the trajectories.  One has \qq \theta(t,\Nr)\ =\ 
\overline{\theta(s,\Nx_{_{t,\Nr}}(s))\m +\m\smallint\limits_s^t
  f(\tau,\Nx_{_{t,\Nr}}(\tau))\, d\tau}
\label{sc1}
\qqq for the tracer and \qq
\rho(t,\Nr)&=&\overline{\rho(s,\Nx_{_{t,\Nr}}(s))
  \,{_{\da(\Nx_{t,\Nr}(s))}\over^{\da(\Nr)}}\m +\m\smallint\limits_s^t
  f(\tau,\Nx_{_{t,\Nr}}(\tau))
  \,{_{\da(\Nx_{t,\Nr}(\tau))}\over^{\da(\Nr)}}\, d\tau}
\label{sc2}
\qqq for the density, where \qq {d\Nx_{_{t,\Nr}}\over
  ds}=v(s,\Nx_{_{t,\Nr}})\m +\m\sqrt{2\kappa}\,{d{\bf{\beta}} \over
  ds}\m,\qquad\Nx_{_{t,\Nr}}(t)=\Nr\m,
\label{tr1}
\qqq with the overbar in Eqs.\,\,(\ref{sc1}) and (\ref{sc2}) denoting
the average over the $d$-dimensional Brownian motions
${\bf{\beta}}(s)$.  \vskip 0.3cm

Clearly, the statistics of the scalar fields reflects the statistics
of the Lagrangian trajectories or, for $\kappa>0$, of their
perturbations by the Brownian motions. In particular, it will be
important to look at the probability distribution functions (p.d.f.'s)
of the difference $\Nr'$ of time $s$ positions of two Lagrangian
trajectories (perturbed by independent Brownian motions if
$\kappa>0$), given their time $t$ positions $\Nr_1$ and $\Nr_2$, \qq
P_2^{t,s}(\Nr_1-\Nr_2,\m\Nr')\,=\,
\langle\,\delta(\Nr'-\Nx_{_{t,\Nr_{1}}}(s)+\Nx_{_{t,\Nr_{2}}}(s))
\,\rangle\m.
\label{jdf2}
\qqq Note that $P_2^{t,s}$ is normalized to unity with respect to $\Nr'$ 
and the equivalent expressions \qq P_2^{t,s}(\Nr_1-\Nr_2,\m\Nr')&=&
\int\langle\,\delta(\Nr'+\Nr-\Nx_{_{t,{\bf r}_{1}}}(s))
\,\,\delta(\Nr-\Nx_{_{t,{\bf r}_{2}}}(s))\,\rangle\,\,d\Nr \cr
&=&\int\langle\,\delta(\Nr'-\Nx_{_{t,{\bf r}_{1}-{\bf r}}}(s))
\,\,\delta(-\Nx_{_{t,{\bf r}_{2}-{\bf r}}}(s))\,\rangle\,\,d\Nr\m,
\label{alte}
\qqq where the last equality uses the homogeneity of the
velocities\footnote{$\Nx_{t,{\bf r}_i-{\bf r}}+{\bf r}\,$ coincides
with $\Nx_{t,{\bf r}_i}$ in the velocity field shifted in space by
$\Nr$}.  \vskip 0.3cm

In the Kraichnan model, the p.d.f.'s $P_2^{t,s}$ may be easily
computed\footnote{the calculation goes back, essentially, to
  \cite{Kr68}}.  They are given by the heat kernels $\ee^{-|t-s|\m
  M_2^\kappa}(\Nr,\m\Nr')$ of the $2^{\rm nd}$-order elliptic 
differential operators $M_2^{\kappa}\,=\,-\m
d^{\alpha\beta}(\Nr)\m\nabla_{\alpha}\nabla_{\beta}\,
-\,2\kappa\nabla^2$. What this means is that the Lagrangian
trajectories undergo, in their relative motion, an effective diffusion
with the generator $M_2^{\kappa}$, i.e. with a space-dependent
diffusion coefficient proportional to their relative distance to power
$\xi$ (for distances large enough that the contribution of the
$\kappa$-term to $M_2^{\kappa}$ may be neglected). Note that, due to
the stationarity and the time reflection symmetry of the velocity
distribution, \qq P_2^{t,s}(\Nr,\m\Nr')\,=\,P_2^{s,t}(\Nr,\m\Nr')
\label{tre2}
\qqq but that, in general,
$P_2^{t,s}(\Nr,\m\Nr')\not=P_2^{s,t}(\Nr',\m\Nr)$, except for the
incompressible case where the operator $M_2^{\kappa}$ becomes
symmetric.  \vskip 0.3cm

\subsection{Statistics of inter-trajectory distances}

For many purposes, it will be enough to keep track only of the distances
between two Lagrangian trajectories. We shall then restrict the p.d.f.'s 
$P_2^{t,s}$ to the isotropic sector by defining 
\qq
P_2^{t,s}(r,r')\,=\,\int\limits_{SO(d)}P_2^{t,s}(\Lambda\Nr,\m\Nr')\,\m
d\Lambda\, =\,\,\int\limits_{SO(d)}
P_2^{t,s}(\Nr,\m \Lambda\Nr')\,\m d\Lambda\m, 
\qqq 
where $d\Lambda$ stands for the normalized Haar measure on $SO(d)$.
$P_2^{t,s}(r,r')$ is the p.d.f. of the time $s$ distance $r'$ between
two Lagrangian trajectories, given their time $t$ distance $r$.
Clearly, \qq P_2^{t,s}(r,r')\ =\ \ee^{-|t-s|\m M_2^\kappa}(r,r') \qqq
with the operator $M_2^\kappa$ restricted to the isotropic sector. In
the action on rotationally invariant functions, 
\qq 
M_2^\kappa\,=\,-\m
Z\m r^{\xi-a}\da_r r^a\da_r\,-\,2\kappa\, r^{-d+1}\da_r
r^{d-1}\da_r\m,
\label{m2ri}
\qqq 
where 
\qq Z\m=\m\CS^2+\xi\m\CC^2\qquad{\rm and}\qquad
a\,=\,[(d-1+\xi)\m\CS^2-\xi\m\CC^2]\m Z^{-1}\m.
\label{a}
\qqq 
The radial Laplacian which constitutes the $\kappa$-term
of $M_2^\kappa$ should be taken with the Neumann boundary
conditions at $r=0$ since the smooth rotationally invariant functions 
on $\NR^d$ satisfy $\da_rf(0)=0$. This is the term that dominates
at small $r$ and, consequently, we should choose the same
boundary condition\footnote{this corresponds to
the domination of the short distances behavior of the perturbed 
trajectories by the independent Brownian motions with the diffusion 
constants $\kappa$} for the complete operator $M_2^\kappa$.
%, i.e. to the motion of fluid particles that do not stick on contact
The adjoint operator $(M_2^\kappa)^*$ with respect to the $L^2$ 
scalar product $\Vert f\Vert^2=\smallint\limits_0^\infty 
\vert f(r)\vert^2 d\mu_d(r)$, where
$d\mu_d(r)=S_{d-1}\m r^{d-1}\m dr$ with $S_{d-1}$ standing for the
volume of the unit sphere in $d$-dimensions, should be taken with the
adjoint boundary conditions which make the integration by parts
possible.  The diagonalization of $M_2^\kappa$ (in the isotropic
sector), if possible, would then permit to write \qq P_2^{t,s}(r,r')\,=
\,\int\ee^{-|t-s|\m E}\,\phi_E(r)\,\psi_E(r') \,\m d\nu(E)\m,
\label{spd}
\qqq where $\phi_E$ and $\psi_E$ stand for the eigen-functions of the
operators $M_2^\kappa$ and $(M^\kappa_2)^*$, respectively, and
$d\nu(E)$ for the spectral measure.  We could naively expect that the
same picture remains true for $\kappa=0$ when \qq
M_2\,\equiv\,M_2^0\,=\,-Z\m r^{\xi-a}\da_r r^{a}\da_r
\label{m2ri0}
\qqq in the rotationally invariant sector. The problem is that the
principal symbol of the operator $M_2$ vanishes at $r=0$ so
that the operator looses ellipticity there and more
care is required in the treatment of the boundary condition.  
\vskip 0.3cm

We start by a mathematical treatment of the problem whose physics we
shall discuss later. It will be convenient to introduce the new
variable $u=r^{2-\xi\over2}$ and to perform the transformation \qq
(Uf)(u)\,=\,({_{2\m S_{d-1}}\over^{2-\xi}})^{^{1\over2}}
\,u^{{d\over2-\xi}-{1\over2}}\, f(u^{2\over2-\xi})
\label{U}
\qqq mapping unitarily the space of square integrable rotationally
invariant functions on $\NR^d$ to $L^2(\NR_+, du)$.  The
transformation $U$, together with a conjugation by a multiplication
operator, turns $M_2$ into the well known Schr\"{o}dinger operator on
the half-line: \qq N_2\,\equiv\, u^{-c}\, U\,M_2\, U^{-1}\m u^{c}\,
=\,{Z'}\,\m[\m-\da_u^2\, +\,
{_{b^2-{_1\over^4}}\over^{u^{2}}}\m]\m,
\label{n2}
\qqq where 
\qq 
Z'\,=\,{_{(2-\xi)^2}\over^4}\,Z\m,\qquad
b\,=\,{_{1-a}\over^{2-\xi}}\qquad{\rm and}\qquad
c\,=\,b+{_d\over^{2-\xi}}-1\m.
\label{b}
\qqq $N_2$ becomes a positive self-adjoint operator in $L^2(\NR_+)$ if
we specify appropriately the boundary conditions at $u=0$. The theory
of such boundary conditions is a piece of rigorous mathematics
\cite{RS}. It says that for $|b|<1$ there is a one-parameter family
of choices of such conditions, among them two leading to the operators
$N_2^\mp$ with the (generalized) eigen-functions 
\qq
\varphi^\mp_E(u)\,=\,u^{\hf}\, J_{\mp{b}}
(\sqrt{E/Z'}\, u)
\label{egf}
\qqq 
(for $b\not=0$) behaving at $u=0$ as $\CO(u^{\hf\mp b})$,
respectively\footnote{the general boundary conditions are
  $u^{|b|-\hf}\varphi(u)|_{_{u=0}}=\lambda\, u^{1-2|b|}\m\da_u\,
  u^{|b|-\hf}\m\varphi(u)|_{_{u=0}}$ with $0\leq\lambda\leq\infty$}.
We then obtain, fixing the spectral measure by the dimensional
consideration and e.g. the action of $N_2^\mp$ on functions $u^\mu$,
\qq \ee^{-|t-s|\m N^\mp_2}(u,u')\,=\,{_1\over^{2\m Z'}}
\int\limits_0^\infty \ee^{-|t-s|\m E}\,
\varphi^\mp_E(u)\,\m\varphi^\mp_E(u')\,\, dE\m.  \qqq Note that the
flip of the sign of $b$ exchanges $N_2^-$ and $N_2^+$.  Relating the
operators $N_2^\mp$ to $M_2^\mp$ by Eq.\,\,(\ref{n2}), we infer that
\qq 
\ee^{-|t-s|\m M_2^\mp}(r,r')\m=\m{_1\over^{2\m Z'}}
\int\limits_0^\infty \ee^{-|t-s|\m E} \,\m
U^{-1}(u^{c}\varphi^\mp_E)(r)\,\m U^{-1}(u^{-c}\varphi^\mp_E)(r')\,\m
dE\cr =\,{_{1}\over^{(2-\xi)\m Z\m S_{d-1}}}\int\limits_0^\infty
\ee^{-|t-s|\m E}\,\, r^{1-a\over 2} \,\, J_{\mp{b}} (\sqrt{E/Z'}\,
{r^{2-\xi\over2}}) \,\hspace{1.2cm}\cr
\cdot\,\,J_{\mp{b}} (\sqrt{E/Z'}\,
{{r'}^{2-\xi\over2}})
\,\,{r'}^{-d+{3\over2}+{a\over 2}-\xi}\,\,dE\m.\hspace{1.2cm}
\label{expo}
\qqq These are explicit versions of the eigen-function
expansion (\ref{spd}) for $\kappa=0$.  
\vskip 0.3cm

The eigen-functions of $M_2^\mp$ can be read of from the above
formula. They behave, respectively, as $\CO(1)$ and
$\CO(r^{1-a})$ at $r=0$.  This is the first choice that corresponds to
the $\kappa\to0$ limit of the Neumann boundary condition for the
operator $M_2^\kappa$. In Appendix A, we analyze a simpler
problem, where the operator (\ref{m2ri}) is replaced by its $\kappa=0$
version (\ref{m2ri0}) made regular by considering it on the interval
$[r_0,\infty[$, with the Neumann boundary condition at $r_0>0$.  We
show that, for $|b|<1$, this is the operator $M_2^-$ that emerges then
in the limit $r_0\searrow0$. The cutting of the interval at a non-zero
value has a similar effect as the addition of the $\kappa$-term to
$M_2$.  We should then have the relation\footnote{the choices
of operators $M_2$ with the other boundary conditions
would describe the trajectories of particles with a tendency 
to aggregate upon the contact and may also have
applications in advection problems}
\qq
P_2^{t,s}(r,r')\,=\,\ee^{-|t-s|\m M_2^-}(r,r')
\label{pts}
\qqq in the $\kappa\to0$ limit, as long as $|b|<1$.  
\vskip 0.3cm

For $|b|\geq 1$, there is only one way to make $N_2$ into a positive
self-adjoint operator. If $b\leq-1$, it is still the operator $N_2^-$
that survives and the relation (\ref{pts}) still holds in the
$\kappa\to0$ limit. For $b\geq1$, however, i.e. for the
compressibility degree $\wp\geq{d\over\xi^2}$, only the operator
$N_2^+$ survives.  Its eigen-functions $\varphi^+_E$ behave as
$\CO(u^{b+\hf})$ at $u=0$ which corresponds to the $\CO(r^{1-a})$
behavior of the eigen-functions of $M_2^+$.  If we impose the Neumann
boundary condition for $M_2$ at $r=r_0$ then, as we show in Appendix
A, in the limit $r_0\searrow0$, the eigen-functions will still become
proportional to the ones obtained from $\varphi^+_E$, not to those
corresponding to $\varphi^-_E$ as it happens for $b<1$.  The same
effect has to occur if we add and then turn off the diffusivity
$\kappa$. It seems then that the equality
$P_2^{t,s}(r,r')\,=\,\ee^{-|t-s|\m M_2^+}(r,r')$ has to hold in the
$\kappa\to0$ limit when $b\geq1$.  \vskip 0.3cm

There is, however, one catch. A direct calculation, see
Eqs.\,\,(\ref{norm-}) and (\ref{norm+}) of Appendix B, shows that the
expression (\ref{pts}) is normalized to unity with respect to $r'$,
but that 
\qq 
\int\limits_0^\infty\ee^{-|t-s|\m M_2^+}(r,r')
\,\,d\mu_d({r'})\,=\,\gamma({b},\m{_{r^{2-\xi}} \over^{4\m
Z'\m|t-s|}})\,\,\Gamma({b})^{^{-1}}\ <\ 1\m,
\label{nor+}
\qqq where $\gamma(b,x)=\smallint\limits_0^xy^{b-1}\m\ee^{-y}\m dy
={x^b\over b}\,\m{}_{_1}\hspace{-0.03cm}F_{_1}(b,1+b;-x)$ is the
incomplete gamma-function. An alternative, but more instructive, way
to reach the same conclusion is to observe that the time derivative of
$\ee^{-t\m M_2^\mp}(r,r')$ brings down the adjoint of $M_2^{\mp}$
acting on the $r'$ variable so that 
\qq 
{_{d}\over^{dt}}\,\m\int\limits_0^\infty\ee^{-t\m M_2^\mp}(r,r') 
\,\,{r'}^{d-1}\m dr'&=&
Z\int\limits_0^\infty\da_{r'}\m{r'}^a\da_{r'}
\m{r'}^{d-1-a+\xi}\left(\ee^{-t\m M_2^\mp}(r,r')\right)\cr
&=&Z\,{r'}^a\da_{r'} \m{r'}^{d-1-a+\xi}\left(\ee^{-t\m
M_2^\mp}(r,r')\right) \bigg\vert^{_{r'=\infty}}_{^{r'=0}}\m.  
\qqq
The contribution from $r'=\infty$ vanishes. On the other hand,
$\,\ee^{-t\m  M_2^-}(r,r')\propto{r'}^{-d+1+a-\xi}$ for small $r'$ 
whereas $\,\ee^{-t\m  M_2^+}(r,r')\propto{r'}^{-d+2-\xi}\m$, \m with
the errors suppressed by an additional factor ${r'}^{2-\xi}$.
It follows that the contribution from $r'=0$ is zero 
for $M_2^-$ if $1+a-\xi>0$, which is the same condition as 
$b<1$ or $\wp<{d\over\xi^2}$, but it
is finite for $M_2^+$.  
\vskip 0.3cm

The lack of normalization may seem strange since when we add the
diffusivity $\kappa$ and fix the Neumann boundary conditions then, by
a similar argument as for $M_2^-$ above, the normalization is
assured.  The solution of the paradox is that for $b\geq1$, the
$\kappa\to0$ convergence of $\ee^{-t\m M_2^\kappa}(r,r')$ to $\ee^{-t\m
M_2^+}(r,r')$ holds only for $r'\not=0$ and the defect of probability
concentrates at $r'=0$.  For $\kappa=0$, we should then add to
$\ee^{-t\m M_2^+}(r,r')$ a delta-function term carrying the missing
probability.  \vskip 0.3cm

We infer this way that \qq \lim\limits_{\kappa\to0}\ P^{t,s}_2(r,r')\ 
=\ \cases{\hbox to 7.9cm{$\ee^{-|t-s|\m M^-_2}(r,r')$\hfill} {\rm
    for}\quad\wp<{d\over\xi^2}\m,\cr\cr \ee^{-|t-s|\m M^+_2}(r,r')\ +\ 
  [\m1\,-\, \gamma({b},\m{_{r^{2-\xi}}\over^{
      4\m Z'\m|t-s|}})\,\,\Gamma({b})^{^{-1}}]\,\m \delta(\Nr')\cr
  \hbox to 7.9cm{{}\hfill}{\rm for}\quad\wp\geq {d\over\xi^2}\m.}
\label{summ}
\qqq In the both cases, the p.d.f.'s $P_2^{t,s}$ satisfy the evolution
equation \qq \da_t\, P_2^{t,s}(r,r')\ =\ \mp\m M_2\, P_2^{t,s}(r,r')
\label{kzee}
\qqq where $M_2$, given by Eq.\,\,(\ref{m2ri0}), acts on the
$r$-variable and the sign $\mp$ corresponds to $t{_{>}\atop^{<}}s$.  They
also have the composition property: \qq \int\limits_0^\infty
P_2^{t,t'}(r,r')\,\m P_2^{t',t''}(r',r'') \m\, d\mu_d(r')\ =\ 
P_2^{t,t''}(r,r'') \nonumber \qqq if $t<t'<t''$ or $t>t'>t''$.  \vskip
0.3cm

It is instructive to note the long time behavior of the averaged
powers of the distance between the Lagrangian trajectories.  As
follows from Eqs.\,\,(\ref{norm-}) and (\ref{norm+}) of Appendix B,
for $\mu>0$, \qq \int\limits_0^\infty
P^{t,s}_2(r,r')\,\,{r'}^{\mu}\,d\mu_d({r'}) \ \ \sim\ \ \cases{\hbox
  to 3.1cm{$\vert t -s\vert^{^{\mu\over2-\xi}}$ \hfill}\qquad{\rm
    for}\quad\wp<{d\over\xi^2}\m,\cr \hbox to 3.1cm{$\vert
    t-s\vert^{^{{\mu\over2-\xi}-b}}\, r^{1-a} $\hfill}\qquad{\rm
    for}\quad\wp\geq{d\over\xi^2}\m.}\quad
\label{ltmb}
\qqq \vskip 0.3cm

\subsection{Fully developed turbulence versus chaos}

The two cases $\wp<{d\over\xi^2}$ and $\wp\geq{d\over\xi^2}$
correspond to two physically very different regimes of the Kraichnan
model.  Let us first notice a completely different typical behavior of
Lagrangian trajectories in the two cases.  In the regime
$\wp<{d\over\xi^2}$, which includes the incompressible case
$\CC^2=0$ studied extensively before, see \cite{nice} and the
references therein, the p.d.f.'s $P_2^{t,s}(r,r')$ possess a
non-singular limit\footnote{a more detailed information on how this
  limit is attained is given by Eq.\,\,(\ref{A12}) in Appendix B}: \qq
\lim\limits_{r\to0}\,\, P_2^{t,s}(r,r')\,=\,
{_{2-\xi}\over^{S_{d-1}\m\Gamma(1-{b})\,\, (4\m Z'\m \vert t
    -s\vert)^{1-{b}}}}\,\, r'^{-d+1+a-\xi}\,\,
\ee^{-{{r'}^{2-\xi}\over 4\m Z'\m\vert t-s\vert}}\m.
\label{toz-}
\qqq It follows that, when the time $t$ distance of the Lagrangian
trajectories tends to zero, the probability to find a non-zero
distance between the trajectories at time $s\not=t$ stays equal to
unity: {\bf infinitesimally close trajectories separate in finite time}.
This signals the "fuzzyness" of the Lagrangian trajectories
\cite{slowm,nice} forming a stochastic Markov process already 
in a fixed typical realization of the velocity field,
with the transition probabilities of the process propagating
weak solutions of the passive scalar equation $\da_t\theta
+\Nv\cdot\nabla\theta=0$ \cite{lejan}. Such appearance 
of stochasticity at the fundamental level seems to be an
essential characteristic of fully developed turbulence in the
incompressible or weakly compressible fluids. It is due to the
roughness of typical turbulent velocities which are only H\"{o}lder
continuous with exponent ${\xi\over 2}<1$ (in the limit of infinite
Rynolds number $Re$).  One should stress an important difference
between this type of stochasticity and the stochasticity of chaotic
behaviors.  In chaotic systems, the trajectories are uniquely
determined by the initial conditions but depend sensitively on the
latter.  The nearby trajectories separate exponentially in time at
the rate given by a positive Lyapunov exponent. The exponential
separation implies, however, that infinitesimally close
trajectories take infinite time to separate. This type of behavior
is observed in flows with intermediate Reynold numbers but for large
Reynolds numbers it occurs only within the very short dissipative
range which disappears in the limit $Re=\infty$.  In the Kraichnan
model, the exponential separation of trajectories characterizes the
$\xi\to 2$ limit of the fuzzy regime $\wp<{d\over\xi^2}$
\cite{kol,slowm}.  \vskip 0.3cm

{\bf In short}, fully developed turbulence and chaos, are two different
things although both lead to stochastic behaviors. In a metaphoric
sense, the difference between the two occurrences of stochasticity is
as between that, more fundamental, in quantum mechanics and that in
statistical mechanics imposed by an imperfect knowledge of microscopic
states.  \vskip 0.3cm

\subsection{Shock wave regime}

Let us discuss now the second regime of our system with
$\wp\geq{d\over\xi^2}$. In that interval,
$\lim\limits_{r\to0}\,\ee^{-t\m M_2^+}(r,r')\,=\,0\,$ and \qq
\lim\limits_{r\to0}\,\, P_2^{t,s}(r,r')\,=\,\delta(\Nr')\m,
\label{toz+}
\qqq see the second of Eqs.\,\,(\ref{summ}).  Here the uniqueness of
the Lagrangian trajectories passing at time $t$ by a given point is
preserved (in probability).  However, with positive probability
tending to one with $\vert t-s\vert \to\infty$, two trajectories at
non-zero distance at time $t$, collapse by time $s$ to zero distance,
as signaled by the presence of the term proportional to $\delta(\Nr')$
in $P^{t,s}_2(r,r')$. The {\bf collapse of trajectories} exhibits the
trapping effect of compressible velocities. A similar behavior is
known from the Burgers equation describing compressible velocities
whose Lagrangian trajectories are trapped by shocks and then move
along with them. The trapping effect is also signaled by the decrease
with time of the averages of low powers of the distance between
trajectories ($<1-a$), see Eq.\,\,(\ref{ltmb}), Note, however, that
the averages of higher powers still increase with time
signaling the presence of large deviations from the typical
behavior.  \vskip 0.3cm

Due to the inequalities $0\leq\wp\leq1$, the second regime,
characterized by the collapse of trajectories, is present only if
$\xi^2\geq d$, i.e. for $d\leq 4$. Its limiting case with $\xi=2$ and
$d\leq 4$ was first discovered and extensively discussed in
\cite{CKV1} and \cite{CKV2}. It appears when the largest Lyapunov 
exponent of (spatially) smooth velocity fields becomes negative. 
\vskip 0.3cm

\nsection{Advection of a tracer: direct versus inverse cascade}

\subsection{Free decay} 

Let us study now the time $t$ correlation functions of the scalar
$\theta$ whose evolution is given by Eq.\,\,(\ref{ps}). Assume first
that we are given a homogeneous and isotropic distribution of
$\theta$ at time zero and we follow its free decay at later times. From 
Eqs.\,\,(\ref{sc00}) and (\ref{alte}) we infer that \qq
F^\theta_2(t,r)&\equiv& \langle\,\theta(t,\Nr)\,\theta(t,{\bf
  0})\,\rangle\ = \int\limits_0^\infty
P^{t,0}_2(r,r')\,\,F_2^\theta(0,r')\,\m d\mu_d({r'})\m.
\label{ind}  
\qqq In particular, to calculate the mean "energy" density
$e_{_\theta}(t)\equiv\langle\m\hf\m\theta(t,\Nr)^2\rangle= \hf\m
F_2^\theta(t,\bf{0})$, the separation $r$ should be taken equal to
zero.  For $\wp<{d\over\xi^2}$, the limit $P^{t,0}_2(0,r')$ is a
regular positive function and it stays such even for $\kappa=0$, see
Eq.\,\,(\ref{toz-}). Since $F_2^\theta(0,r')\leq F_2^\theta(0,0)$ as a
Fourier transform of a positive measure, it follows that the total
energy diminishes with time: $e_{_\theta}(t)<e_{_\theta}(0)$.  On the
other hand, for $\wp\geq{d\over\xi^2}$,
$P^{t,0}_2(0,r')=\delta(\Nr')$, see Eq.\,\,(\ref{toz+}), and the total
energy is conserved: $e_{_\theta}(t)=e_{_\theta}(0)$.  The loss of
energy in the regime $\wp<{d\over\xi^2}$ is not due to
compressibility\footnote{in temporally decorrelated velocity
fields, the mean energy $e_{_\theta}$ is conserved also
in compressible flows, in the absence of 
forcing and diffusion}, but to the non-uniqueness of the Lagrangian
trajectories responsible for the persistence of the short-distance
dissipation in the $\kappa\to0$ limit. As is well known this
dissipative anomaly accompanies the direct cascade of energy
towards shorter and shorter scales in the (nearly) incompressible
flows. On the other hand, in the strongly compressible regime
$\wp\geq{d\over\xi^2}$, the scalar $\theta$ is transported
along unique trajectories and its energy is conserved in mean. The
short distance dissipative effects disappear in the limit
$\kappa\to0$: there is no dissipative anomaly and no direct cascade of
energy. As we shall see, the energy injected by the source of $\theta$
is transferred instead to longer and longer scales in an inverse
cascade process.  \vskip 0.3cm

\subsection{Forced state for weak compressibility}

The direction of the energy cascade may be best observed if we keep
injecting the energy into the system at a constant rate.  Let us then
consider the advection of the tracer in the presence of stationary
forcing. From Eqs.\,\,(\ref{sc1}) and (\ref{fc}), assuming that 
$\theta$ vanishes at time zero, we obtain, \qq
F^\theta_2(t,r)=\langle\,\smallint\limits_0^t f(s,\Nx_{t,\Nr}(s))\,ds
\smallint\limits_0^t f(s',\Nx_{t,0}(s'))\,ds'\,\rangle=
\int\limits_0^tds\int\limits_0^\infty P^{t,s}_2(r,r') \chi({r'})\,
d\mu_d({r'}),\quad
\label{dif}
\qqq which is a solution of the evolution equation 
\qq 
\da_t
F^\theta_2\,=\,-M_2^{\kappa}\m F^\theta_2\,+\,\chi\quad
\label{diffe}
\qqq 
with the operator $M_2^{\kappa}$ given by Eq.\,\,(\ref{m2ri}).
\vskip 0.3cm

When $\wp<{_{d-2+\xi}\over^{2\m\xi}}$ (i.e. for $a>1$ or $b<0$), which
implies that we are in the weakly compressible phase with
$\wp<{d\over\xi^2}$, and for $\kappa=0$, 
\qq
F^\theta_2(t,r)&=&\ \smallint\limits_0^tds\smallint\limits_0^\infty
\ee^{-s\m M_2^-}(r,r')\,\m\chi({r'}) \,\m d\mu_d({r'})\ \ 
\mathop{\rightarrow}\limits_{t\to\infty}\ \ 
\smallint\limits_0^\infty(M_2^-)^{-1}(r,r')\,\m\chi({r'}) \,\m
d\mu_d({r'})\cr
&=&\m{_1\over^{(a-1)\m
    Z}}\m\smallint\limits_r^\infty {r'}^{1-\xi}\,\chi({r'})\,
dr'\,+\,{_1\over^{(a-1)\m Z}}\,
r^{1-a}\smallint\limits_0^r{r'}^{a-\xi}\,\chi({r'})\, dr'
\ \,\equiv\ \,F_2^\theta(r)\m,
\label{35}
\qqq see Eq.\,\,(\ref{a24}) of Appendix C. Thus for
$\wp<{_{d-2+\xi}\over^{2\m\xi}}$, when $t\to\infty$, the 2-point
function of $\theta$ attains a stationary limit $F^\theta_2(r)$
with a finite mean energy density $e_{_\theta}=\langle\m\hf\m\theta^2\rangle
=\hf\m F^\theta_2(0)$. 
The corresponding stationary 2-point structure function is \qq
&&S^\theta_2(r)\ =\ \langle\,(\theta(\Nr)-\theta({\bf 0}))^2\m\rangle
\,=\,2\m(F^\theta_2(0)-F^\theta_2(r))\,=\,{_2\over^Z}\m
\smallint\limits_0^r\zeta^{-a}\,d\zeta\smallint\limits_0^\zeta
{r'}^{a-\xi}\,\chi({r'})\, dr'\cr\cr&&\ \cong\ \cases{
%{_{2\m\chi(0)}\over^{(2-\xi)\m(d\CS^2-\xi^2\CC^2)}}
  \hbox to 10.2cm{${_{2\m\chi(0)}\over^{(2-\xi)(1+a-\xi)\m Z}}
    \,r^{2-\xi}$\hfill}{\rm for\ }r\ {\rm small}\m,\cr \hbox to
  10.2cm{${_2\over^{(a-1)\m Z}}\,
    \smallint\limits_0^\infty{r'}^{1-\xi}\,\chi({r'})\,
    dr'\,-\,{_2\over^{(a-1)\m Z}}\, r^{-(a-1)}\,
    \smallint\limits_0^\infty{r'}^{a-\xi}\,\chi({r'})\,
    dr'$\hfill}{\rm for\ }r\ {\rm large}\m.}
\label{stf}
\qqq Thus $S^\theta_2(r)$ exhibits a normal scaling at $r$ much
smaller than the injection scale $L$ whereas at $r\gg L$ the approach
to the asymptotic value $2\langle\,\theta^2\,\rangle$ is controlled by
the scaling zero mode $r^{1-a}$ of the operator $M_2$.
In Appendix D, we give the explicit form of
the stationary 2-point function $F^\theta_2$ in the presence of
positive diffusivity $\kappa$.  \vskip 0.3cm

\subsection{Dissipative anomaly}

Let us recall how the dissipative anomaly manifests itself in this
regime. The stationary 2-point function of the tracer solves the
stationary version of Eq.\,\,(\ref{diffe}).  When we let in the latter
$r\to0$ for positive $\kappa$, only the contribution of the dissipative
term in $M_2^\kappa$ survives and we obtain the energy balance equation
$\epsilon_{_\theta}\,\equiv\,\kappa\,\langle\m(\nabla\theta)^2\rangle
\,=\,\hf\m\chi(0)$, i.e. the equality of the mean dissipation rate
$\epsilon_{_\theta}$ and the mean energy injection rate
$\hf\m\chi(0)$.  Taking first $\kappa\to0$ and $r\to0$ next, instead,
we obtain the analytic expression of the dissipative anomaly: \qq
\lim\limits_{\kappa\to0}\,\,\epsilon_{_\theta}\,
=\,\hf\,\lim\limits_{r\to0}\,M_2\,
\lim\limits_{\kappa\to0}\,F^\theta_2(r) \,=\,\hf\m\chi(0)\m.
\label{da}
\qqq Thus, in spite of the explicit factor $\kappa$ in its definition,
the mean dissipation rate does not vanish in the limit $\kappa\to0$,
which explains the name: anomaly.  \vskip 0.3cm

For ${_{d-2+\xi}\over^{2\m\xi}}<\wp<{d\over\xi^2}$ (i.e.
for $0<b<1$, see Eq.\,\,(\ref{b})), the 2-point function
$F_2^\theta(t,r)$, still given for $\kappa=0$ by the left
hand side of the relation (\ref{35}),   
%\qq F^\theta_2(t,r)\ =\ 
%\smallint\limits_0^tds\smallint\limits_0^\infty \ee^{-s\m
%  M_2^-}(r,r')\,\m\chi({r'}) \,\m d\mu_d({r'})
%\label{dint}
%\qqq 
diverges with time as $t^{b}$. More exactly, as we show in Appendix C, 
it is the expression
\qq
&&F^\theta_2(t,r)\ -\ {_{(4\m Z')^{b}}\over^{(1-a)\m Z\,
    \Gamma({1-b})}}\,\, t^{b}\,\m
\smallint\limits_0^\infty\m{r'}^{a-\xi}\,\chi({r'})\, dr'
%\cr
%&&\mathop{\rightarrow}\limits_{t\to\infty}\ \, -\m{_1\over^{(1-a)\m
%    Z}}\m\smallint\limits_r^\infty {r'}^{1-\xi}\,\chi({r'})\,
%dr'\,-\,{_1\over^{(1-a)\m Z}}\,
%r^{1-a}\smallint\limits_0^r{r'}^{a-\xi}\,\chi({r'})\, dr'\m.
\label{toA}
\qqq that tends to the right hand side of the relation (\ref{35}).
Finally, for $\wp={_{d-2+\xi}\over^{2\m\xi}}$, there is a
constant contribution to $F^\theta_2(t,r)$ logarithmically divergent
in time. For 
${_{d-2+\xi}\over^{2\m\xi}}\leq\wp<{d\over\xi^2}$
%${_{2\m\xi}\over^{d+2-\xi}}\m\CC^2\geq\CS^2>{_{\xi^2}\over^d}\m\CC^2$, 
the system still dissipates energy
at short distances with the rate $\epsilon_{_\theta}$ that becomes
equal to the injection rate asymptotically in time, but it also builds
up the energy $e_{_\theta}(t)$ in the constant mode with the rate 
decreasing as $t^{-(1-b)}$. Note that in spite of the divergence 
of the 2-point correlation function, the 2-point structure function 
of the tracer still converges as $t\to\infty$ to a stationary form 
given by Eq.\,\,(\ref{stf}). Now, however, $S^\theta_2(r)$
is dominated for large $r$ by the growing zero mode $\propto r^{1-a}$
of $M_2$.  
\vskip 0.5cm

\subsection{Forced state for strong compressibility}

Let us discuss now what happens under steady forcing in the strongly 
compressible regime $\wp\geq{d\over\xi^2}$ (i.e. for $1+a-\xi\leq0$ 
or $b\geq1$). Here the 2-point function (\ref{dif}), which still
evolves according to Eq.\,\,(\ref{diffe}), is for $\kappa=0$ given
by the relation
\qq
F^\theta_2(t,r)&=&\smallint\limits_0^tds\smallint\limits_0^\infty
\ee^{-s\m M_2^+}(r,r')\,\m\chi({r'}) \,\m d\mu_d({r'})\cr
&+&\chi(0)\,\smallint\limits_0^t [\m 1\,-\,
\gamma({b},\m{_{r^{2-\xi}}\over^{4\m Z'\m s}})\,\,
\Gamma({b})^{^{-1}}]\m\,ds\m.
\label{2pf}
\qqq 
When $t\to\infty$, the first term on the right hand side tends to 
\qq 
&&\hspace{-0.7cm}\smallint\limits_0^\infty(M^+_2)^{-1}(r,r')\,\chi({r'})\, 
d\mu_d({r'})\cr
&&\hspace{1.5cm}={_1\over^{(1-a)\m Z}}\m \smallint\limits_0^r{r'}^{1-\xi}
\,\chi({r'})\,dr'+{_1\over^{(1-a)\m Z}}\, 
r^{1-a}\smallint\limits_r^\infty
{r'}^{a-\xi}\,\chi({r'})\,dr'\quad\label{gf0}\\
&&\hspace{1.5cm}\cong\ \cases{\hbox to 7.4cm{$
%-{_{\chi(0)}\over^{(2-\xi)(d\CS^2-\xi^2\CC^2)}}\, r^{2-\xi}$\hfill}
    -\m{_{\chi(0)}\over^{(2-\xi)(1+a-\xi)\m Z}}\, r^{2-\xi}
    \,-\,{_{\smallint\limits_0^\infty {r'}^{1+a-\xi}\,\chi'({r'})\,
        dr'} \over^{(1-a)(1+a-\xi)\m Z}}\, r^{1-a}\m $\hfill}
  \quad{\rm for\ }r\ {\rm small}\m,\cr\cr
%\hbox to 5.5cm{${_1\over^{2\xi\m\CC^2-(d-2+\xi)\CS^2}}\m
  \hbox to 7.4cm{${_1\over^{(1-a)\m Z}}\m
    \smallint\limits_0^\infty{r'}^{1-\xi}\,\chi({r'})\, dr'
    $\hfill}\quad{\rm for\ }r\ {\rm large}\m,}
\label{gf}
\qqq where the asymptotic expressions hold for $1+a-\xi<0$, i.e.
for $\wp>{d\over\xi^2}$.
On the other hand \qq
{_{\chi(0)}\over^{\Gamma({b})}}\,\smallint\limits_0^t
\gamma({b},\m{_{r^{2-\xi}}\over ^{4\m Z'\m(t-s)}})\m\,ds\ 
\ \mathop{\rightarrow}\limits_{t\to\infty}
\ \ -\m{_{\chi(0)}\over^{(2-\xi)(1+a-\xi)\m Z}}\,\m r^{2-\xi}\m,
\label{tin}
\qqq except for $\wp={d\over\xi^2}$ when it diverges as
${\chi(0)\,r^{2-\xi}\over4\m Z'}\,\ln t$.  Hence, for
$\wp>{d\over\xi^2}$, the quantity $F^\theta_2(t,r)\,-\,\chi(0)\m t$
converges when $t\to\infty$ and the limit is proportional
to the zero mode $r^{1-a}$ of $M_2$ for small $r$ 
(up to $\CO(r^{4-\xi})$ terms). As we see, the energy injected into 
the system by the external source is accumulating in the constant mode
with the constant rate equal to the injection rate $\hf\m\chi(0)$.
The dissipative anomaly is absent in this phase. Indeed, 
\qq
\epsilon_{_\theta}\,=\,\hf\,\lim\limits_{r\to0}\,\lim\limits_{t\to\infty}
\,M_2\, F^\theta_2(t,r)\ =\ 0 
\qqq 
and the same is true at finite times since $F^\theta_2(t,r)$
becomes proportional to the zero modes of $M_2$ at short distances.
These are clear signals of the inverse cascade of energy towards large
distances, identified already in the $\xi\to2$ limit of the
$\wp\geq{d\over\xi^2}$ regime in \cite{CKV1,CKV2}.
\vskip 0.3cm

The 2-point structure function \qq
S^\theta_2(t,r)\,=\,2\m(F_2^\theta(t,0)-F_2^\theta(t,r))\,=\, 2\m\chi(0)\m
t\,-\,2\m F_2^\theta(t,r)\m,
\label{2psf}
\qqq which satisfies the evolution equation \qq \da_t S^\theta_2\ =\ 
-\m M_2\m S^\theta_2\,+\, 2\m(\chi(0)-\chi)\m,
\label{2pfe}
\qqq reaches for $\wp>{d\over\xi^2}$ the stationary limit
%\qq S^\theta_2(r)&=&-\m{_2\over^{(1-a)\m Z}}
%\smallint\limits_0^\infty{r'}^{1-\xi}\,\chi({r'})\, dr'\m
%+\m{_2\over^Z}\m\smallint\limits_r^\infty\zeta^{-a}\, d\zeta
%\smallint\limits_{\zeta}^\infty{r'}^{a-\xi}\, \chi({r'})\,dr'\m
%-\m{_{2\,\chi(0)}\over^{(2-\xi)(1+a-\xi)\m Z}}\,r^{2-\xi}\cr
%&&\hspace{3cm}\cong\cases{\hbox to
%  5cm{${_{2\m\smallint\limits_0^\infty {r'}^{1+a-\xi}\,\chi'({r'})\,
%        dr'} \over^{(1-a)(1+a-\xi)\m Z}}\, r^{1-a}\m$\hfill} {\rm for\ 
%    }r\ {\rm small}\m,\cr\cr \hbox to
%  5cm{$-\m{_{2\,\chi(0)}\over^{(2-\xi)(1+a-\xi)\m Z}} \,\m
%    r^{2-\xi}$\hfill}{\rm for\ }r\ {\rm large}\m,}
%\label{s2c}
%\qqq 
whereas it diverges logarithmically in time for
$\wp={d\over\xi^2}$.  Note that it is now at large $r$ that
$S^\theta_2(r)$ scales normally $\propto r^{2-\xi}$ and at small $r$ 
that it becomes proportional to the zero mode $r^{1-a}$ of $M_2$.  
\vskip 0.3cm

\nsection{Intermittency of the direct cascade}

The higher correlation functions of the convected scalars involve
simultaneous statistics of several Lagrangian trajectories.
To probe deeper into the statistical properties of the trajectories, 
it is convenient to consider the joint p.d.f.'s
$P^{t,s}_{_N}(\Nr_1,\dots,\Nr_{_N};\m\Nr'_1,\dots,\Nr'_{_N})$ of the
time $s$ differences of the positions $\Nr'_1,\dots\Nr'_{N}$ of $N$
Lagrangian trajectories passing at time $t$ through points
$\Nr_1,\dots,\Nr_{_N}$. In the notation of Section 2, \qq
P^{t,s}_{_N}(\Nr_1,\dots,\Nr_{_N};\m\Nr'_1,\dots,\Nr'_{_N}) \ =\ 
\int\langle\,\prod\limits_{n=1}^N
\delta(\Nr'_n-\Nx_{_{t,\Nr_n}}(s)+\Nr)\,\rangle\,\, d\Nr\m.  
\qqq 
Clearly, the functions $P_N^{t,s}(\un{\Nr};\m\un{\Nr'})$
are translation-invariant separately in the variables $\un{\Nr}=
(\Nr_1,\dots,\Nr_{_N})$ and in $\un{\Nr'}=(\Nr'_1,\dots,\Nr'_{_N})$.
In the Kraichnan model, the p.d.f.'s $P^{t,s}_{_N}$ are again
given by heat kernels of degree two differential operators
\cite{schlad} 
\qq 
P^{t,s}_{_N}(\un{\Nr};\m\un{\Nr'}) \ =\ \ee^{-\vert
  t-s\vert\m M^\kappa_{_N}}(\un{\Nr};\un{\Nr'})\m, 
\qqq 
where the operators 
\qq
M^\kappa_{_N}\ =\ \sum\limits_{1\leq n<m\leq N} 
d^{\alpha\beta}(\Nr_n-\Nr_m)
\,\nabla_{r_n^\alpha}\nabla_{r_m^\beta}\ -\ \kappa \sum\limits_{1\leq
  n\leq N}\nabla_{\Nr_n}^2
\qqq 
should be restricted to the
translation-invariant sector, which enforces the separate
translation-invariance of their heat kernels. The relations 
$P^{t,s}_{_N}(\un{\Nr};\m\un{\Nr'}) =P^{s,t}_{_N}(\un{\Nr};\m\un{\Nr'})$ 
generalize Eq.\,\,(\ref{tre2}). As for $N=2$,
$P^{t,s}_{_N}(\un{\Nr};\m\un{\Nr'}) =P^{s,t}_{_N}(\un{\Nr'};\m\un{\Nr})$ 
only in the incompressible case. 
\vskip 0.3cm

Strong with the lesson we learned for two trajectories, 
we expect completely different behavior of the p.d.f's
$P^{t,s}_{_N}(\un{\Nr};\m\un{\Nr'})$ for $\Nr_n$'s close to each other
in the two phases, resulting in different short-distance
statistics of convected quantities. Let us start by discussing 
the weakly compressible case $\wp<{d\over\xi^2}$.  Here 
we have little to add to the incompressible story, 
see e.g. \cite{slowm,nice}. We expect the limit 
$\lim\limits_{\un{\Nr}\to\un{\bf 0}}\ P^{t,s}_{_N}(\un{\Nr};
\m\un{\Nr'})\ \equiv\ P^{t,s}_{_N}(\un{\bf 0};\m\un{\Nr'})$ to exist
and to be a continuous function (except, possibly at
$\un{\Nr'}=\un{\bf 0}$) decaying with $\vert t-s\vert$ and at large
distances, just as for $P^{t,s}_{2}$, see Eq.\,\,(\ref{toz-}). More
exactly, we expect \cite{slowm} an asymptotic expansion generalizing
the expansion (\ref{A12}) of Appendix B for $P^{t,s}_2$: 
\qq
P^{t,s}_{_N}(\lambda\un{\Nr};\m\un{\Nr'})\ \ =\ \ 
\sum\limits_{{i\atop j=0,1,\dots}}\lambda^{\sigma_i+(2-\xi)j}
\,\,\phi_{i,j}(\un{\Nr})\,\,\overline{\psi_{i,j}(\vert t-s\vert,
  \m\un{\Nr'})} 
\qqq 
for $\lambda$ small, where $\phi_{i,0}$ are
scaling zero modes of the operator $M^0_{_N}\equiv M_{_N}$ 
with scaling dimensions
$\sigma_i\geq0$ and $\phi_{i,p}$ are "slow modes", of scaling
dimension $\sigma_i+(2-\xi)j$, satisfying the descent equations
$M_{_N}\phi_{i,j}= \phi_{i,j-1}$. The constant zero mode
$\phi_{0,0}=1$ (corresponding to $\overline{\psi_{0,0}}=P_{_N}(\un{\bf
  0}; \m\,\cdot\m\,)$) gives the main contribution for small $\lambda$, 
but drops out if
we consider combinations of $P^{t,s}_{_N}
(\lambda\un{\Nr};\m\un{\Nr'})$ with different configurations
$\un{\Nr}$ which eliminate the terms that do not depend on all
(differences) of $\Nr_n$'s.  Such combinations are dominated by the
zero modes depending on all $\Nr_n$'s. For small $\xi$, there is one
such zero mode $\phi_{i_0,0}$ for each even $N$.  A perturbative
calculation of its scaling dimension done as in \cite{BKG}, where the
incompressible case was treated, gives \qq \sigma_{i_0}\ =\ 
N\,-\,(\m{_N\over^2}\m+\m{_{N(N-2)\m(1+2\wp)}
  \over^{2(d+2)}}\m)\,\xi\ +\ \CO(\xi^2)\ \equiv \ 
{_N\over^2}(2-\xi)\,+\,\Delta_{_N}^\theta\m.
\label{Ncf}
\qqq \vskip 0.3cm

In the absence of forcing, the $N$-point
correlation functions $\,F^\theta_{_N}(t,\un{\Nr})=\langle\m\prod
\limits_{n=1}^N\theta(t,\Nr_n)\m\rangle\,$ of the tracer
are propagated by the p.d.f.'s $P^{t,s}_{_N}\m$:
\qq
F^\theta_{_N}(t,\un{\Nr})\ =\ \int P^{t,s}_{_N}(\un{\Nr};
\m\un{\Nr'})\,\,F^\theta_{_N}(s,\un{\Nr})\,\,d'\un{\Nr}'
\label{fdn}
\qqq
where $d'\un{\Nr'}\equiv d\Nr'_2\cdots d\Nr'_{_N}$,
compare to Eq.\,\,(\ref{ind}).
In the presence of forcing, $F_{_N}^\theta$ obey
recursive evolution equations \cite{ss,schlad}. 
%\qq \da_t\,
%F^\theta_{_N}(t,\un{\Nr})\ =\ -M_{_N}\,F^\theta_{_N}(t,\un{\Nr}) \ +\ 
%\sum\limits_{1\leq n<m\leq N}\hspace{-0.1cm}
%F^\theta_{_{N-2}}(t,\m\Nr_1,\mathop{\dots\dots} \limits_{\hat{n}\ \ 
%  \hat{m}},\m\Nr_{_N})\ \chi({\vert\Nr_n-\Nr_m\vert})\m.\quad
%\label{Ncfe}
%\qqq
If $F^\theta_{_N}$ vanish at time zero then 
the odd correlation functions
vanish at all times and the even ones may be computed iteratively:
\qq 
F^\theta_{_N}(t,\un{\Nr})\ =\ 
\int\limits_0^t ds\int P^{t,s}_{_N}(\un{\Nr};\m\un{\Nr'})
\,\sum\limits_{n<m}F^\theta_{_{N-2}}(s,\m\Nr'_1,
\mathop{\dots\dots}\limits_{\hat{n}\ \ \hat{m}}, \Nr'_{_N})\ 
\chi({\vert \Nr'_n-\Nr'_m\vert})\ d'\un{\Nr'}\m.
\label{Nc1}
\qqq 
We expect that for small $\xi$ or weak compressibility, 
$F^\theta_{_N}(t,\un{\Nr})$ tend at large times 
to the stationary correlation functions
$F^\theta_{_N}(\un{\Nr})$ whose parts depending on all $\Nr_n$'s are
dominated at short distances by the zero modes of $M_{_N}$. In
particular, this scenario leads to the anomalous scaling of the
$N$-point structure functions $S^\theta_{_N}(r)=\langle\m(\theta(\Nr)
-\theta({\bf 0}))^N\m\rangle$ which pick up the contributions to
$F^\theta_{_N}$ dependent on all $\Nr_n$'s.  Naively, one could
expect that $S^\theta_{_N}(r)$ scale for small $r$ with powers
${N\over2}(2-\xi)$, i.e. ${N\over2}$ times the 2-point function
exponent.  Instead, they scale with smaller exponents, which signals
the small scale intermittency: 
\qq 
S^\theta_{_N}(r)\ \ \propto\ \ 
r^{(2-\xi){_N\over^2} +\Delta^\theta_{_N}}
\label{ansc}
\qqq with the anomalous (part of the) exponent $\Delta^\theta_{_N}$ 
given for small $\xi$ by \qq \Delta^\theta_{_N}\ =\ 
-\,{_{N(N-2)\m(1+2\m\wp)} \over^{2(d+2)}}\,\xi\ +\ 
\CO(\xi^2)\m,
\label{aex}
\qqq see Eq.\,\,(\ref{Ncf}). We infer that the {\bf direct cascade is
  intermittent}.  \vskip 0.3cm

\nsection{Absence of intermittency in the inverse cascade}

\subsection{Higher structure functions of the tracer}

For $\wp\geq{d\over\xi^2}$, i.e.  in the strongly
compressible phase, we expect a completely different behavior of the
p.d.f.'s $P^{t,s}_{_N}(\un{\Nr}; \m\un{\Nr'})$ when the points $\Nr_n$
become close to each other. The (differences of) Lagrangian
trajectories in a fixed realizations of the velocity field are
uniquely determined in this phase if we specify their time $t$
positions. The p.d.f.'s for $N$ trajectories should then reduce to 
those of $N-1$ trajectories if we let the time $t$ positions of two 
of them coincide: \qq \lim\limits_{\Nr_{_N}\to\Nr_{_{N-1}}} \,
P^{t,s}_{_N}(\un{\Nr};\m\un{\Nr'}) \ =\ 
P^{t,s}_{_{N-1}}(\Nr_1,\dots,\Nr_{_{N-1}};\m\Nr'_1,\dots,
\Nr'_{_{N-1}})\,\,\delta(\Nr'_{_{N-1}}-\Nr'_{_N})\m.
\label{contr}
\qqq 
Applying this relation $N$ times, we infer, that 
$\,P^{t,s}_{_N}(\Nr,\dots,\Nr;\m\un{\Nr'})\ =\ \prod\limits_{n=2}^N
\delta(\Nr'_1-\Nr'_n)\m.$ \,Since the p.d.f.'s $P_{_N}^{t,s}$ propagate 
the $N$-point functions of the tracer in the free decay, 
see Eq.\,\,(\ref{fdn}), it follows that, in the strongly compressible 
phase, such a decay preserves all the higher mean quantities 
$\langle\m\theta(t,\Nr)^N\m\rangle=F^\theta_N(t,\Nr,\dots,\Nr)$.
In the presence of forcing, however, all these quantities 
are pumped by the source. Indeed, Eq.\,\,(\ref{Nc1}) implies now 
that
\qq
\langle\m\theta(t,\Nr)^N\m\rangle\,=\,{_{N(N-1)}\over^2}\,\chi(0)
\smallint\limits_0^t\langle\m\theta(s,\Nr)^{N-2}\m\rangle\,\m ds\m,
\label{thn}
\qqq
from which it follows that, for even $N$, $\,\langle\m\theta(t,\Nr)^N\m
\rangle=(N-1)!!\,(\chi(0)\m t)^{^{N\over2}}$.
\vskip 0.3cm

The relation (\ref{contr}) permits also to calculate effectively the higher
structure functions $S^\theta_{_N}(t,r)$ in the strongly compressible
phase. We prove in Appendix E that for $N$ even, 
\qq
S^\theta_{_N}(t,r)\ =\ N(N-1)\,\smallint\limits_0^tds
\smallint\limits_0^\infty P_2^{t,s}(r,r')\,\,S^\theta_{_{N-2}}(s,r')
\,\,(\chi(0)-\chi({r'}))\,\,d\mu_d({r'})\m.
\label{rrsn}
\qqq Note that $S^\theta_{_N}$ satisfies the evolution equation 
\qq
\da_t\m S^\theta_{_N}\ =\ -\m M_2\m S^\theta_{_N}\,+\,
N(N-1)\,S^\theta_{_{N-2}}\,\m(\chi(0)-\chi)\m.
\label{rrde}
\qqq 
This is the same equation that would have been obtained directly 
from Eq.\,\,(\ref{ps}) neglecting the viscous term and averaging 
with respect to the Gaussian fields $\Nv$ and $f$, e.g. 
by integration by parts \cite{UF}. 
The situation should be contrasted with that in the weakly
compressible case where the evolution equations for the structure
functions do not close due to the dissipative anomaly 
which adds to Eq.\,\,(\ref{rrde}) terms that are not directly
expressible by the structure functions \cite{Kr94}, see also \cite{BKG}.  
\vskip 0.3cm

Substituting into Eq.\,\,(\ref{rrsn}) the expression (\ref{summ}) for
$P^{t,s}_2$ in the strongly compressible phase, we obtain \qq
S^\theta_{_N}(t,r)\ =\ 
N(N-1)\,\smallint\limits_0^tds\smallint\limits_0^\infty \ee^{-\vert
  t-s\vert\m M_2^+}(r,r')
\,\,S^\theta_{_{N-2}}(s,r')\,\,(\chi(0)-\chi({r'}))
\,\,d\mu_d({r'})\m.\quad
\label{taf}
\qqq The above formula implies, by induction, that $S^\theta_{_N}$ 
are positive functions (no surprise), growing in time. Suppose that
$S^\theta_{_{N-2}}(t,r)$ reaches a stationary form
$S^\theta_{_{N-2}}(r)$ which behaves proportionally to the zero mode
$r^{1-a}$ for small $r$ and which exhibits the normal scaling $\propto
r^{({_N\over^2}-1)(2-\xi)}$ for large $r$ ($S^\theta_2$ behaves this
way for $\wp>{d\over\xi^2}$, i.e. for $b>1$).  Then \qq
\smallint\limits_0^tds\smallint\limits_0^\infty \ee^{-\vert t-s\vert\m
  M_2^+}(r,r')
\,\,S^\theta_{_{N-2}}(s,r')\,\,\chi({r'})\,\,d\mu_d({r'}) \qqq
converges when $t\to\infty$ to a function bounded by \qq
\smallint\limits_0^\infty \m(M_2^+)^{-1}(r,r')
\,\,S^\theta_{_{N-2}}(r')\,\,\chi({r'})\,\,d\mu_d({r'})\m, \qqq which
behaves as $r^{1-a}$ for small $r$ and as a constant for large $r$,
compare to the estimate (\ref{gf}).  On the other hand, the dominant
contribution to the $\chi(0)$ term in Eq.\,\,(\ref{taf}) is
proportional to \qq \smallint\limits_0^tds\smallint\limits_0^\infty
\ee^{-s\m M_2^+}(r,r')\,\,{r'}^{({_N\over^2}-1) (2-\xi)}\,\,
d\mu_d({r'})\m.
\label{mcon}
\qqq Since, by Eq.\,\,(\ref{norm+}) of Appendix B,
$\smallint\limits_0^\infty\ee^{-s\m M_2^+}(r,r')
\,\,{r'}^{({_N\over^2}-1)(2-\xi)}\,\, d\mu_d({r'})$ vanishes at $s=0$
and behaves as $s^{{_N\over^2}-1-b}\m r^{1-a}$ for large $s$, we infer
that the integral (\ref{mcon}) stabilizes when $t\to\infty$ only if
\qq {_N\over^2}\,<\,b\,=\,{_{1-a}\over^{2-\xi}}\m.
\label{stri}
\qqq This condition becomes more and more stringent with increasing
$N$. If it is not satisfied, then the contribution (\ref{mcon}), and,
consequently, $S^\theta_{_N}(t,r)$, diverge as $t^{{_N\over^2}-b}\m
r^{1-a}$. If it is satisfied, the contribution (\ref{mcon}) reaches a
limit when $t\to\infty$ which is proportional to
$r^{{_N\over^2}(2-\xi)}$. It then dominates for large $r$ the
stationary $N$-point structure function $S^\theta_{_N}(r)$ which for
small $r$ behaves as $r^{1-a}$, reproducing our inductive assumptions.
\vskip 0.3cm

{\bf Summarizing}: The even $N$-point structure functions become
stationary at long times only if the conditions (\ref{stri}) are
satisfied and they exhibit then the normal scaling at distances much
larger than the injection scale $L$, i.e.  in the inverse energy
cascade. At the distances much shorter than $L$, however, the existing
stationary structure functions are very intermittent: they scale with
the fixed power $1-a$.  \vskip 0.3cm

\subsection{Generating function and p.d.f. of scalar differences}

It is convenient to introduce the generating function for the
structure functions of the scalar defined by \qq
\CZ^\theta(\lambda,t,r)\ =\ \langle\,\ee^{\m i\m\lambda
  \m(\theta(t,\Nr)-\theta(t,{\bf 0}))}\m\rangle\ =\ 
\sum\limits_{n=0}^\infty{_{(-1)^n\m\lambda^{2n}}\over^{(2n)!}}  \,\,
S^\theta_{2n}(t,r)\,.
\label{gfsd}
\qqq We shall take $\lambda$ real. Note that the evolution equation
Eq.\,\,(\ref{rrde}) implies that \qq \da_t\m \CZ^\theta\ =\ 
-\m[M_2\m+\m\lambda^2\m(\chi(0)-\chi)]\, \CZ^\theta\m.
\label{gfee}
\qqq At time zero, $\CZ^\theta\equiv 1$. Since $M_2+\lambda^2
(\chi(0)-\chi)$ has a similar boundary condition problem at $r=0$ as
$M_2$, one should be careful writing down the solution of the
parabolic equation (\ref{gfee}). It is not difficult to see
that $\CZ^\theta$ is given by a Feynman-Kac type formula:
\qq
\CZ^\theta(\lambda,t,r)\ =\ E_{_r}\Big(\ee^{-\lambda^2
\smallint\limits_0^t(\chi(0)-\chi(r(s)))\, ds}\Big),
\label{cz}
\qqq
where $E_{_r}$ is the expectation value w.r.t. the Markov process
$r(s),\ s\geq 0$, with transition probabilities
$P_2^{t,s}(r,r')$, starting at time zero at $r$. Due to the
delta-function term in the transition probabilities, almost all 
realizations $r(s)$ of the process arrive at finite time at $r=0$ 
and then do not move. Note that $\CZ^\theta(0,t,r) 
=\CZ^\theta(\lambda,t,0)=1$ and that $\CZ^\theta(\lambda,t,r)$ 
decreases in time. Moreover, $\CZ^\theta(\lambda,s,r)$ for $s\geq t$ 
is bounded below by an expectation similar to that of Eq.\,\,(\ref{cz})
but with the additional restriction that $r(t)=0$ (and hence that
$r(s')=0$ for all $s'\geq t$).  The latter is positive since the
probability that $r(t)=0$ is non-zero (it even tends to one when
$t\to\infty$). Thus a non-trivial limit $\lim\limits_{t\to\infty}\,
\CZ^\theta(\lambda,t,r)\equiv\CZ^\theta(\lambda,r)$ exists. It 
satisfies the stationary version of Eq.\,\,(\ref{gfee}): 
\qq
[M_2\m+\m\lambda^2\m(\chi(0)-\chi)]\,\CZ^\theta\ =\ 0\m.  
\qqq 
In particular, for $r$ large, for which we may drop $\chi(r)$,
$\CZ^\theta(\lambda,r)$ is an eigen-function of the operator $M_2$
given by Eq.\,\,(\ref{m2ri0}) with the eigen-value $-\lambda^2\chi(0)$.
This permits to find the analytic form of the generating
function $\CZ^\theta(\lambda,r)$ in this regime, a rather rare 
situation in the study of models of turbulence. We have 
\qq 
\CZ^\theta(\lambda,r)\ \cong\ {_{2^{1-b}}\over^{\Gamma(b)}}
\m\left(\sqrt{\chi(0)/Z'}\,\vert\lambda\vert \m
r^{_{2-\xi}\over^2}\right)^{\hspace{-0.03cm}b}\,K_b
({\sqrt{\chi(0)/Z'}} \,\vert\lambda\vert\,
r^{\xi-2\over2})\,\equiv\,\CZ^\theta_{\rm sc} (\lambda^2\m
r^{2-\xi})\m.\ \quad 
\qqq 
The Bessel function $K_b(z)$ decreases exponentially 
at infinity. We have chosen the normalization so that
$\CZ^\theta(0,r)=1$. Since $z^b\m K_b(z)$ has an expansion around
$z=0$ in terms of $z^{2n}$ and $z^{2b+2n}$ with $n\geq0$, it is
$N$-times differentiable at zero only if ${N\over2}<b$.  Not
surprisingly, this is the same condition that we met above as the
requirement for the existence of stationary limits of the structure
functions $S^\theta_{_N}(t,r)$.  \vskip 0.3cm

{\bf In short}: in the strongly compressible phase
$\wp\geq{d\over\xi^2}$, the generating function
$\CZ^\theta(\lambda,t,r)$ has a stationary limit $\CZ^\theta(\lambda,r)$
which for large distances takes the scaling form $\CZ^\theta_{\rm sc}
(\lambda^2 r^{2-\xi})$. Although $\CZ^\theta(\lambda,r)$ is non-Gaussian 
and even not a smooth function of $\lambda$, its normal scaling in 
the large $r$ regime, responsible for the normal scaling of the
existing structure functions, signals the {\bf suppression 
of intermittency in the inverse cascade}.  
\vskip 0.3cm

The same behavior may be seen in the Fourier transform of the
generating function $\CZ^\theta(\lambda,t,r)$ giving the p.d.f. of
scalar differences: 
\qq
\label{distribuzione}
\CP^\theta(t,\vartheta,r)\ \equiv\ \langle\,\m\delta(\vartheta
-\theta(t,\Nr)+\theta(t,{\bf 0}))\,\rangle\ =\ {_1\over^{2\pi}}\,
\int\ee^{-i\m\lambda\m\vartheta}\,\CZ^\theta(\lambda,t,r)\,
d\lambda\m.  
\qqq 
The $t\to\infty$ limit $\CP^\theta(\vartheta,r)$ of the finite-time
p.d.f. satisfies the partial-differential
equation $\,[\m M_2\,-\,(\chi(0)-\chi)\,\da_{\vartheta}^2\m]\,
\CP^\theta\,=\, 0\m.$ \,For $r$ large, the latter reduces to the
ordinary differential equation 
\qq
\da_x\m[\m(\chi(0)\m+\m Z'\m x^2)\,\da_x\,+
\,(2b+1)\m Z'\m x\m]\m\, p^\theta\ =\ 0\m, 
\qqq 
where $\,\CP^\theta(\vartheta,r)=r^{-{_{2-\xi}\over^2}}\m
p^\theta(r^{-{_{2-\xi}\over^2}}\m\vartheta)\m.$ \,The normalized,
smooth at $x=0$ solution is 
\qq 
p^\theta(x)\ =\ {_{\sqrt{Z'}\,\chi(0)^b\, \Gamma(2b)}
\over^{2^{2b-1}\m\Gamma(b)^2}}\ [\m\chi(0)\m
+\m Z'\, x^2\m]^{^{-b-{1\over2}}}\m.
\label{ppp}
\qqq 
It is the Fourier transform of the generating function
$\CZ^\theta_{sc}(\lambda^2)$. Note that the condition ${N\over2}<b$
becomes now the condition for the existence of the $N$-th moment of
$p^\theta$. The slow decay of $p^\theta(x)$ at infinity renders most
of the moments divergent.

\nsection{Infrared cutoffs and the inverse cascade}

As shown in the previous section, in the strongly compressible phase, 
the asymptotic behavior of the scalar $\theta$ is quasi-stationary: 
due to the excitation of larger and larger scales, observables might 
or might not reach a stationary form. It is therefore of fundamental 
interest to analyze how the inverse cascade properties are affected 
by the presence of an infrared cutoff at the very large scales. 
This has also a practical importance as such cutoffs are always present 
in concrete situations in one form or another \cite{Tab,SY,Betal,Bisk}. 
The simplest modification of the dynamics that introduces an infrared 
cutoff is to add to (\ref{ps}) a friction term: 
\qq 
\da_t
\theta+\Nv\cdot\nabla\theta+\alpha\,\theta-\kappa\nabla^2\theta=f\m,
\label{friction}
\qqq 
where $\alpha$ is a positive constant. We shall be interested in
studying the limit when $\alpha\to 0$. For flows smooth in space and
$\delta$-correlated in time, the case considered in \cite{MC}, the
advection and the friction terms have the same dimensions. For
non-smooth flows, this is not the case and friction and advection
balance at the friction scale
$\eta_{_f}\sim\CO\left(\alpha^{-1/(2-\xi)}\right)$ which becomes arbitrarily
large when $\alpha$ tends to zero. Roughly speaking, advection and
friction dominate at scales much smaller and much larger than $\eta_{_f}$,
respectively.  The hierarchy of scales is therefore the mirror image
of the one for the direct cascade\,: the energy is injected at the
integral scale $L$, transferred upwards by the advection term
in the convective range of the inverse cascade and finally extracted 
from the system at very large scales. We are interested in the influence 
of the infrared cutoff scale on the inverse cascade convective range 
properties and we shall therefore assume that $\wp\geq{d\over\xi^2}$ 
throughout this section.

\vskip 0.3cm Heuristically, it is {\it a priori} quite clear that
the friction will make the system reach a stationary state. 
Specifically, the friction term in Eq.\,\,(\ref{friction}) is
simply taken into account by remarking that the field
$\tilde\theta(t,\Nr)=\exp(\alpha\,t)\m\theta(t,\Nr)$ satisfies the
equation (\ref{ps}) with a forcing $\ee^{\alpha\m t}\m f(t,\Nr)$. We can
then carry over the Lagrangian trajectory statistics from the previous
Sections and we just have to calculate the averages with the
appropriate weights. For example, the recursive equation (\ref{Nc1})
for the $N$-point function in the presence of forcing becomes
\qq  
F^\theta_{_N}(t,\un{\Nr})\m = 
\int\limits_0^t ds\int\ee^{-(t-s)\m N\alpha}\,
P^{t,s}_{_N}(\un{\Nr};\m\un{\Nr'})
\sum\limits_{n<m}F^\theta_{_{N-2}}(s,\m\Nr'_1,
\mathop{\dots\dots}\limits_{\hat{n}\ \ \hat{m}}, \Nr'_{_N})\ 
\chi({\vert \Nr'_n-\Nr'_m\vert})\ d'\un{\Nr'}.\ \quad
\label{Nc1a}
\qqq 
%\qq
%\label{due}
%\langle\,\theta(t,\Nr)\,\theta(t,\Nr')\,\rangle=\int_0^t\,ds\,\int_0^\infty
%e^{-2\alpha(t-s)}\,P_2^{t,s}(r,r')\chi(r')\,d\mu_d(r') .  
%\qqq 
Similarly, the expressions (\ref{thn}) and (\ref{rrsn})
for $\langle\m\theta(t,\Nr)^N\m\rangle$ and $S^\theta_{_N}(t,r)$
are modified by inclusion of the factor $\ee^{-(t-s)\m N\alpha}$
under the time integrals. This renders them convergent 
in the limit $t\to\infty$, in contrast with the $\alpha=0$ case.
As a result, 
$\langle\m\theta(t,\Nr)^N\m\rangle$ and $S^\theta_{_N}(t,r)$
reach when $t\to\infty$ the limits that are the solutions
of the stationary versions of the evolution equations
\qq
&&\da_t\m\langle\m\theta^N\rangle\,=\,{_{N(N-1)}\over^2}\,\chi(0)\,
\langle\m\theta^{N-2}\rangle\,-\, N\alpha\,\langle\m\theta^N\rangle\m,
\label{etha}\\
&&\da_t\m S^\theta_{_N}\ =\ -\m M_2\m S^\theta_{_N}\,+\,
N(N-1)\,S^\theta_{_{N-2}}\,\m(\chi(0)-\chi)\,-\,N\alpha\, S^\theta_N\m.
\label{rrda}
\qqq 
We obtain then in the stationary state: $\langle\m\theta(\Nr)^N\rangle=
(N-1)!!\,({\chi(0)\over2\alpha})^{^{N\over2}}$ and 
\qq
\label{soluzione}
S^\theta_{_N}(r)\,=\,\left({_{\chi(0)}\over^\alpha}
\right)^{^{\hspace{-0.1cm}{N\over 2}}}(N-1)!!
\,\m\Big[1+{_N\over^{2^b\m\Gamma(b)}}\sum_{k=1}^{N\over 2}{_{(-1)^k}\over^k}
\left({{}_{{N\over 2}-1}\atop{}^{k-1}}\right) 
z_k^b\,K_b\left(z_k\right)\Big], 
\qqq where the variables $z_k$ are defined as 
\qq
\label{etaf}
z_k\equiv 2\m k^{\hf}\m({_r\over^{\eta_{_f}}})^{^{2-\xi\over2}}\qquad
{\rm with}\qquad
\eta_{_f}\equiv ({_{2\m Z'}\over^{\alpha}})^{^{1\over2-\xi}}
\qqq 
being the friction scale. For $r\gg\eta_{_f}$, all the Bessel 
functions tend to zero and $S^\theta_{_N}$ 
reaches the constant asymptotic value $2^{^{{N\over 2}}}\hspace{-0.05cm}
(N-1)!!\,({_{\chi(0)}\over^{2\alpha}})
^{^{N\over 2}}$  which agrees with the stationary value of
$2^{^{{N\over 2}}}\hspace{-0.05cm}\langle\,\theta^N\,\rangle$. 
The expansion of $K_b$ for small 
arguments gives the following dominant behaviors in
the inverse cascade convective range $L\ll r\ll \eta_{_f}$: 
\qq
\label{risposta}
S^\theta_{_N}(r)\ \cong\ \left\{ \begin{array}{ll}
    c_1\,\,r^{^{{N\over2}(2-\xi)}} & \mbox{if\quad$b>{_N\over^2}$}\m,\\
    c_2\,\,r^{^{{N\over2}(2-\xi)}}\log\left({\eta_{_f}\over r}\right)
    & \mbox{if\quad$b={_N\over^2}$}\m,\\
    c_3\,\,r^{^{{N\over2}(2-\xi)}} \left({\eta_{_f}\over r}\right)
    ^{(2-\xi)({N\over2}-b)}&
    \mbox{if\quad$b<{N\over2}$}\m,
                        \end{array}\right\}
\qqq 
where the constants $c_i$ are given by 
\qq
\label{costanti}
                        \begin{array}{ll}
                c_1=\left(\chi(0)\over 4\m Z'\right){N!\over(N/2)!}
                {\Gamma(b-N/2)\over \Gamma(b)}\m,\quad\qquad 
                c_2=\left(\chi(0)\over 4\m Z'\right){N!\over(N/2)!}
                {2-\xi \over \Gamma(b)}\m,\\
                                        \\
                c_3=\left(\chi(0)\over 4\m Z'\right){N!\over(N/2-1)!}
                {\Gamma(1-b)\over \Gamma(1+b)}
                \left(-\sum\limits_{k=1}^{N/2}{(-1)^k\over k}\left(
                {{N/2-1}\atop{k-1}}\right)k^b\right).
                        \end{array}
                        \qqq
The threshold $b={N\over2}$ in Eq.\,\,(\ref{risposta}) is the
same as in the inequality (\ref{stri}), discriminating the
moments that do not converge at large times in the absence of
friction. The converging moments are not
modified by the presence of friction.
Conversely, those that were diverging are now
tending to finite values but they pick an
anomalous scaling behavior in the cutoff scale
$\eta_{_f}$. Note that the moments with ${N\over2}>b$
scale all with the scaling exponent $1-a$.

%---------------------------
\begin{figure}
\begin{center}
\vspace{-0.6cm}
\mbox{\hspace{0.0cm}\psfig{file=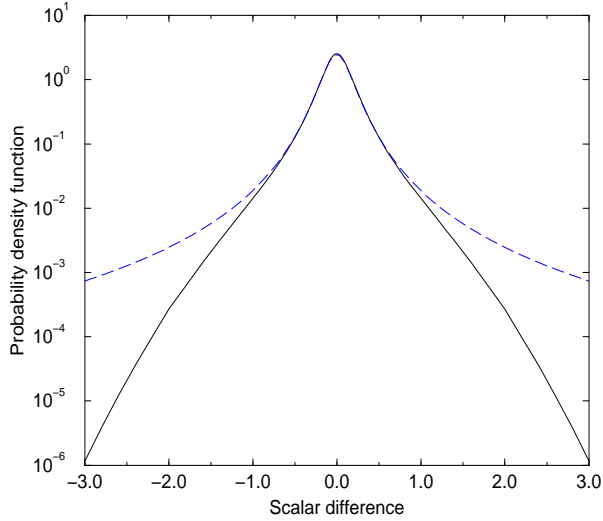,height=8cm,width=9cm}}
\end{center}
\vspace{-0.6cm}
\caption{{\it The probability distribution function of scalar differences with 
and without friction (solid and dashed lines). The specific parameters 
are $d=1$, $\xi=1$, ${\cal S}^2={\cal C}^2=\hf$, $\chi(0)=1$, $\alpha=2$, 
and $r=0.01$.}}
\label{fig1}
\end{figure}
%---------------------------

\vskip 0.3cm

It is interesting to look at this saturation
from the point of view of the p.d.f.
$\CP^\theta(\vartheta,r)$, defined
as in the relation (\ref{distribuzione}).  The equation for
$\CP^\theta$ can be derived by the same
procedure as in the previous section. Its
stationary version reads 
\qq
\label{proba}
-M_2\CP^\theta\,+\,\alpha\m\,\da_{\vartheta}\left(\vartheta\CP^\theta\right)
+\,(\chi(0)-\chi)\,\da_{\vartheta}^2\, \CP^\theta\ =\ 0\m.
\qqq 
For the scales $r\gg L$ of interest to us here, $\chi$ can be neglected
with respect to $\chi(0)$.
The relevant informations on the p.d.f. $\CP^\theta$ are conveniently 
extracted by expanding the function $\CP^\theta$ in a series 
\qq
\label{espansione}
\CP^\theta(\vartheta,r)=\sqrt{{_\alpha\over^{2\pi\m\chi(0)}}}\,
\ee^{-{\alpha\vartheta^2\over 2\chi(0)}}\sum_{k=0}^\infty d_{2k}(r)\,
H_{2k}\left(\sqrt{{_\alpha\over^{2\chi(0)}}}\,\vartheta\right) 
\qqq 
of the Hermite polynomials $H_{2k}$. The coefficients $d_{2k}$ can be
obtained by plugging the expansion (\ref{espansione}) into 
Eq.\,\,(\ref{proba}) and using well-known properties of Hermite 
functions. One obtains 
\qq
\label{coefficienti}
d_0(r)=1\m,\qquad d_{2k}(r)={_{(-1)^k}\over^{\Gamma(b)}}{_1\over^{k!\,\m
2^{b-1+2k}}}\,z_k^b\m\,K_b(z_k)\m, 
\qqq where $z_k$ is defined in Eq.\,\,(\ref{etaf}). At
scales $r\gg\eta_{_f}$, friction dominates and the p.d.f. tends to a
Gaussian form. In the inverse cascade convective range $L\ll r\ll\eta_{_f}$, 
the solution (\ref{ppp}) remains valid as long as $\vartheta^2\ll
{\chi(0)\over\alpha}$, while the power-law tails are cut by an
exponential fall off. The situation is exemplified in Fig.~1. The
scaling behavior (\ref{risposta}) of the structure functions is then easy
to grasp: for ${N\over2}<b$, the dominant contribution comes from the scale
invariant part of the p.d.f., resulting in the normal scaling, whereas for
${N\over2}\geq b$, the leading contribution comes from the region
around $\theta^2={\chi(0)\over\alpha}$, with the tails cut out by friction.
The dominant behavior can be captured by simply calculating the moments 
with the scale-invariant p.d.f. (\ref{ppp}) cut at 
$\vartheta^2={\chi(0)\over\alpha}$.

\nsection{Advection of a density}

The time $t\,$ 2-point function of the scalar $\rho$, whose advection is
governed by Eq.\,\,(\ref{psb}), may be studied similarly as for the
tracer $\theta$, see \cite{russ,mazverg}.  For the free decay
of the 2-point function, we obtain \qq F^\rho_2(t,r)\equiv
\langle\,\rho(t,\Nr)\,\rho(t,{\bf 0})\,\rangle\ = \ 
\int\limits_0^\infty P^{0,t}_2(r',r) \,\,F_2^\rho(0,r')\,\m
d\mu_d({r'})\m,
\label{ind1}
\qqq where we have used Eqs.\,\,(\ref{sc00}) and (\ref{alte}), compare
to Eq.\,\,(\ref{ind}). The evolution (\ref{psb}) of the scalar $\rho$
preserves the total mass $\int\hspace{-0.08cm}\rho(t,\Nr) \m d\Nr$. 
As a consequence,
the mean ``total mass squared'' per unit volume
$m^2_\rho(t)\equiv\int\langle\m\rho(t,\Nr)\m \rho(t,{\bf
  0})\m\rangle\m d\Nr=\int\limits_0^\infty F^\rho_2(t,r) \, d\mu_d(r)$
does not change in time in both phases.  \vskip 0.3cm

In the presence of the stationary forcing, the 2-point function of
$\rho$ computed with the use of Eqs.(\ref{sc2}) and (\ref{alte}) becomes 
\qq 
F^\rho_2(t,r)\ =\ 
%\int\limits_0^tds\int\langle\,
%\delta(\Nr-\Nx_{_{s,{\bf r}'}}(t)) \,\m\delta(-\Nx_{_{s,{\bf
%      r}''}}(t))\,\rangle
%\,\m\chi(\vert\Nr'-\Nr''\vert)\,\,d\Nr'\,d\Nr''\cr
%&=&
\int\limits_0^tds\int\limits_0^\infty P^{s,t}_2(r',r)
\,\m\chi({r'})\,\m d\mu_d({r'})\m,
\label{dif1}
\qqq 
if $\rho=0$ at time zero. It evolves according to the equation
\qq 
\da_t F^\rho_2\,=\,-(M_2^{\kappa})^*\m F^\theta_2\,+\,\chi\m,
\label{diffd}
\qqq 
i.e. similarly to $F^\theta_2$, see Eq.\,\,(\ref{diffe}), but with 
$M_2^{\kappa}$ exchanged for its adjoint
$(M_2^{\kappa})^*$, a signature of the duality between
the two scalars noticed before.   
\vskip 0.3cm

For $\wp<{_{d-2+\xi}\over^{2\m\xi}}$, the 2-point function
$F^\rho_2(t,r)$ attains a stationary form given by
Eq.\,\,(\ref{2pkar}) of Appendix D for $\kappa>0$ and reducing for
$\kappa=0$ to the expression 
\qq F^\rho_2(r)\ =\ {_1\over^{(a-1)\m
    Z}}\, r^{-d+1+a-\xi}\smallint\limits_r^\infty
{r'}^{d-a}\,\m\chi(r')\,\m dr'\m+\m{_1\over^{(a-1)\m Z}}
\,r^{-d+2-\xi}\smallint\limits_0^r{r'}^{d-1}\,\m\chi(r')
\,\m dr'\m.\ \quad
\label{36}
\qqq
%\\
%\cr &&\ \cong\ \ \cases{\hbox to 6.3cm{$-\m{_1\over^{(1-a)\m Z}}\,
%    r^{-d+1+a-\xi}\smallint\limits_0^\infty{r'}^{d-a}\,
%    \chi({r'})\,dr'$\hfill}\ \quad{\rm for\ }r\ {\rm small}\m,\cr
%  \hbox to 6.3cm{$ -\m{_1\over^{(1-a)\m
%        Z}}\,r^{-d+2-\xi}\smallint\limits_0^\infty
%    {r'}^{d-1}\,\m\chi(r')\,\m dr'$\hfill} \ \quad{\rm for\ }r\ {\rm
%    large}\m.}\quad
%\label{361}
%\qqq 
In particular, $F_2^\rho(r)$ becomes proportional to $r^{-d+1+a-\xi}$
for small $r$ and diverges at $r=0$, except for the
incompressible case when the two scalars coincide.  The small $r$
behavior agrees with the result of \cite{russ} and, in one dimension,
with that of \cite{mazverg}. For large distances $r$, 
the function $\,F_2^\rho(r)$ is proportional to $r^{-d+2-\xi}$. 
In the upper interval 
${_{d-2+\xi}\over^{2\m\xi}}<\wp<{_d\over^{\xi^2}}$
of the weakly compressible phase, it is 
$$F^\rho_2(t,r)\m-\m {_{(4\m Z')^{b}}\over^{(1-a)\m
Z\,\Gamma({1-b})}}\,\,t^{b}\,r^{-d+1+a-\xi}\smallint\limits_0^\infty
{r'}^{d-1}\,\chi(r')\,\m dr'$$
that reaches the stationary limit still given for $\kappa=0$ 
by the right hand side of Eq.\,\,(\ref{36}). Thus the 2-point function
$F_2^\rho(t,r)$ is pumped now into the zero mode $r^{-d+1+a-\xi}$
of the operator $M_2^*$.
\vskip 0.3cm

The higher order correlation functions of $\rho$,
$\,F^\rho_N(t,\un{\Nr})\m\equiv\m\langle\m\prod\limits_{n=1}^N
\rho(t,\Nr_n)\m\rangle$, \,are expected to converge for long times to
a stationary form for sufficiently small $\xi$ and/or compressibility
and to be dominated at short distances by a scaling zero mode $\psi_0$
of the operator $M_N^*$, \qq \psi_0(\un{\Nr})\ =\ 
1\,-\,d\,{\wp}\, \xi\sum\limits_{1\leq n<m\leq N}
\ln{\vert\Nr_n-\Nr_m\vert}\ +\ \CO(\xi^2)\m.
\label{zmrho}
\qqq The zero mode $\psi_0$ becomes equal to 1 when $\xi\to0$.  The
scaling dimension of $\psi_0$ may be easily calculated to the first
order in $\xi$ by applying the dilation generator to the left hand
side of Eq.\,\,(\ref{zmrho}).  It is equal to \qq -\m{_{N(N-1)\m d}
\over^{2}}\m \wp\,\xi\ +\CO(\xi^2) \ \equiv\ 
{_N\over^2}(2-\xi)\,+\,\Delta^\rho_{_N}\m, \qqq which again agrees
with the result 
$$\Delta^\rho_{_N}\,=\, -\m N\,
+\,{_{N\m(1-(N-1)\,\wp\m d)}\over^{2}}\,\xi\ 
+\ \CO(\xi^2)$$
of \cite{russ} and with the exact result $\,\Delta^\rho_2=a-1-d\,$
obtained above. Note the singular behavior of
$\psi_0(\un{\Nr})$ at the origin, at least for small $\xi$.  
\vskip 0.4cm

Finally, in the strongly compressible phase
$\wp\geq{d\over\xi^2}$, where the second of the
expressions (\ref{summ}) has to be used for $P_2^{s,t}$, we obtain 
\qq
F^\rho_2(t,r)&=&\smallint\limits_0^tds\smallint\limits_0^\infty
\ee^{-s\m M^+_2}(r',r)\,\m\chi(r')\,\m d\mu_d(r')\cr &&+\ 
\delta(\Nr)\,\smallint\limits_0^tds\smallint\limits_0^\infty
[\m1\m-\m\gamma(b,\m{_{{r'}^{2-\xi}}\over^{4\m Z'\m
    s}})\,\m\Gamma(b)^{-1}]\,\,\chi(r')\,\m d\mu_d(r')\m.
\qqq 
For large times, $F_2^\rho(t,r)$ is pumped into 
the singular mode proportional to the delta-function
with a constant rate\footnote{the pumping disappears if
$\smallint_0^\infty\chi(r)\m d\mu_d(r)=0$, which is the case
considered in \cite{mazverg,russ}, but even then $F_2^\rho$
picks up a singular contribution in the limit $\kappa\to0$}: 
\qq 
&&F^\rho_2(t,r)\ -\ \delta(\Nr)\,\m
t\,\m\smallint\limits_0^\infty \chi(r')\,\m d\mu_d(r')\ \ 
\mathop{\rightarrow}\limits_{t\to\infty} \ \ {_1\over^{(1-a)\m Z}}\,
r^{-d+1+a-\xi}\smallint\limits_0^r {r'}^{d-a}\,\chi(r')\,\m dr'\cr
&&+\ {_1\over^{(1-a)\m Z}}\,r^{-d+2-\xi}\smallint\limits_r^\infty
{r'}^{d-1}\,\chi(r')\,\m dr'\ +\ {_{S_{d-1}}\over^{(2-\xi)(1+a-\xi)\m
    Z}}\,\delta(\Nr)\smallint\limits_0^\infty{r'}^{d+1-\xi}
\,\m\chi(r')\,\m dr' 
\qqq 
(except for $\wp={d\over\xi^2}$), compare to Eqs.\,\,(\ref{gf0})
and (\ref{tin}). Its non-singular part, however, stabilizes
and becomes proportional to $r^{-d+2-\xi}$ for small $r$ and 
to $r^{-d+1+a-\xi}$ for large $r$. Note the inversion
of the powers as compared with the weakly compressible phase.
%\qq F^{\rho}_{2,\m\rm ns}(r)&=&
%{_1\over^{(1-a)\m Z}}\, r^{-d+1+a-\xi}\smallint\limits_0^r
%{r'}^{d-a}\,\chi(r')\,\m dr'\ +\ {_1\over^{(1-a)\m
%    Z}}\,r^{-d+2-\xi}\smallint\limits_r^\infty
%{r'}^{d-1}\,\chi(r')\,\m dr'\cr \cr &&\ \cong\ \ \cases{\hbox to
%  6.1cm{${_1\over^{(1-a)\m Z}}\,
%    r^{-d+2-\xi}\smallint\limits_0^\infty{r'}^{d-1}\,
%    \chi({r'})\,dr'$\hfill}\ \quad{\rm for\ }r\ {\rm small}\m,\cr
%  \hbox to 6.1cm{$ {_1\over^{(1-a)\m
%        Z}}\,r^{-d+1+a-\xi}\smallint\limits_0^\infty
%    {r'}^{d-a}\,\m\chi(r')\,\m dr'$\hfill} \ \quad{\rm for\ }r\ {\rm
%    large}\m.}\quad
%\label{362}
%\qqq 
\vskip 0.3cm

The mean total mass squared of $\rho$ per unit volume, $m^2_\rho$,
exhibits in the presence of forcing a position-space cascade analogous
to the wavenumber-space cascade of the energy $e_{_\theta}$.  Let us
localize $m^2_\rho$ in space by defining its amount between the radii
$r$ and $R$ as \qq m^2_{\rho;\m r,R}(t)\ =\ \smallint\limits_r^R
F^\rho_2(t,r') \,\m d\mu_d(r')\m.
\label{eloc}
\qqq Integrating the evolution equation (\ref{diffd}) for 
$F^\rho_2$ with the radial
measure $d\mu_d$ from $r$ to $R$, and using the explicit form of
$\,M_2^*=-\m Z\, r^{-d+1}\m\da_r\m r^a\m\da_r\m
r^{d-1-a+\xi}\m,$ \,we obtain the relation
\qq 
\da_t\,m^2_{\rho;\m r,R}(t)\ =\ Z\m S_{d-1}\m {r'}^a\m\da_{r'}\m
{r'}^{d-1-a+\xi}\,F^\rho_2(t,r')\Big\vert_{r}^{R}\ 
+\ \smallint\limits_r^R\chi(r')\,\m d\mu_d(r') 
\qqq 
expressing the local balance of the total mass
squared, provided that we interpret $\smallint\limits_r^R\chi(r')\,\m
d\mu_d(r')$ as the injection rate of $m^2_\rho$ in the radii between
$r$ and $R$ and $$Z\m S_{d-1}\m r^{a}\m\da_r\m r^{d-1-a+\xi}\, 
F^\rho_2(r,t)\,\equiv\,\Phi(r)$$
as the flux of $m^2_\rho$ into the radii $\leq r$.
In the weakly compressible phase $\wp<{_d\over^{\xi^2}}$ 
and in the stationary state, $\Phi(r)=-\m
\smallint\limits_0^r\chi(r')\,\m d\mu_d(r')$ 
so that the flux is constant for $r$ much larger than the injection
scale $L$ and it is directed towards bigger radii. On the other hand, in
the strongly compressible phase $\wp>{_d\over^{\xi^2}}$,
one has the equality $\Phi(r)= \smallint\limits_r^\infty\chi(r')
\,\m d\mu_d(r')$ so that the flux is directed towards smaller 
distances. It eventually feeds into the
singular mode all of $m^2_\rho$ injected by the source.  As we see,
the two phases differ also by the direction of the cascade of the
total mass squared of $\rho$.

\nsection{Conclusions}

We have studied the Gaussian ensemble of compressible $d$-dimensional
fluid velocities decorrelated in time and with spatial behavior
characterized by the fractional H\"{o}lder exponent ${\xi\over2}$. We
have shown that the Lagrangian trajectories of fluid particles in such
an ensemble exhibit very different behavior, depending on the degree
of compressibility $\wp$ of the velocity field. For
$\wp<{_d\over^{\xi^2}}$, i.e. $b$ defined in (\ref{b}) smaller than
unity, the infinitesimally close trajectories separate in finite
time, implying that the dissipation remains finite in the limit
when the molecular diffusivity $\kappa\to 0$ and that 
the energy is transferred towards small scales 
in a direct cascade process. The constancy of the
flux at small scales leads to a normal scaling behavior $r^{2-\xi}$ of
the second order structure function $S^\theta_2(r)$ for $r\ll L$ 
(the typical scale where the energy is injected).  For $b$ negative 
(which includes the incompressible case), as the system evolves, 
the dissipation rate tends to the injection rate rapidly enough 
to ensure that the energy $\langle\theta^2\rangle$ remains finite. 
The non-constant zero mode $r^{(2-\xi)b}$ controls the decay 
of $S^\theta_2(r)$ to its finite asymptotic value 
$2\langle\theta^2\rangle$ at large $r$.  Conversely, for $0\le b<1$, 
the dissipation rate tends to the injection rate very slowly,
$\propto t^{-(1-b)}$, and the energy is thus increasing with time as
$t^b$.  The structure function $S^\theta_2(r)$ grows now at large 
distances as the zero mode $r^{(2-\xi)b}$.  For $b\ge 1$, coinciding
particles do not separate and, in fact, separated particles collapse
in a finite time. The consequences are that the dissipative anomaly is
absent and that the energy is entirely transferred toward larger and 
larger scales in an inverse cascade process.  The threshold $b=1$ 
corresponds to the crossing of the exponents: $(2-\xi)b$ 
of the non-constant zero mode and $2-\xi$ of the constant-flux-of-energy 
solution. The picture is the mirror image of the one for the direct 
cascade, with the first exponent controlling now the small scale 
behavior and the second one appearing at the large scales.  
A sketch of the three possible situations is presented in Fig.~2.

%---------------------------
\begin{figure}
\begin{center}
\vspace{-0.6cm}
\mbox{\hspace{0.0cm}\psfig{file=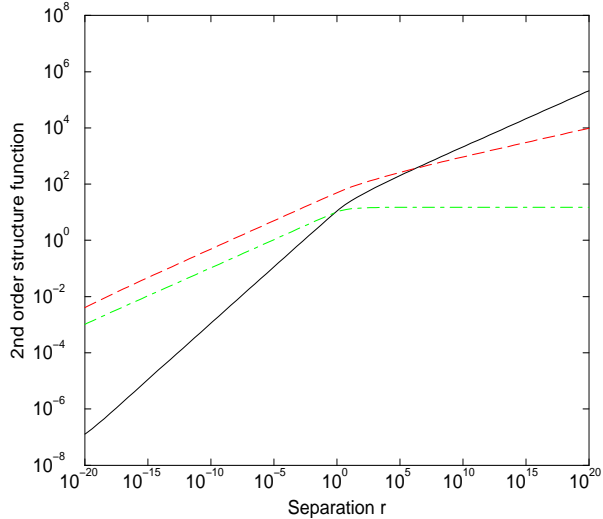,height=8cm,width=9cm}}
\end{center}
\vspace{-0.6cm}
\caption{{\it The second-order structure function $S^\theta_2(r)$ {\it vs} 
  $r$ for $\xi=1.8$ and $d=3$ in the three different regimes $b=-2$
  (dot-dashed line), $b=0.5$ (dashed line) and $b=2$ (solid line).}}
\label{fig2}
\end{figure}
%---------------------------

\vskip 0.3cm

Concerning higher order correlations in the strongly compressible
phase $b\ge 1$, we have shown that the inverse energy cascade is
self-similar, i.e. without intermittency.  The effects of a large
scale friction reintroduce, however, an anomalous scaling of the
structure functions that do not thermalize without friction.
They were exhibited in the explicit expressions for
the p.d.f.'s of the tracer differences. As for
the scalar density, the different behaviors of the Lagrangian
trajectories were shown to result in the inverse or in the direct
cascade of the total mass squared in, respectively, the weakly and the
strongly compressible phase. As explained in Introduction, we expect
the explosive separation of the Lagrangian trajectories and/or their
collapse to persist in more realistic ensembles of fully turbulent
velocities and to play a crucial role in the determination 
of statistical properties of the flows at high
Reynolds numbers.  \vskip 0.5cm

\noindent{\bf Acknowledgements}. \ M. V. was partially supported by 
C.N.R.S. GdR ``M\'{e}canique des Fluides G\'{e}ophysiques et 
Astrophysiques''.
\vskip 0.6cm

\nappendix{A}
%\label{A:one}
\vskip 0.5cm

Let us consider the operator $M_2$ of Eq.\,\,(\ref{m2ri0}) on the
half-line $[r_0,\infty[$, $r_0>0$, with the Neumann boundary condition
$\da_r\m f(r_0)=0$. By the relation (\ref{n2}), this means that we
have to consider the operator $N_2$ on $[u_0,\infty[$ with the
boundary condition \qq \da_u\m u^{b-\hf}\m\varphi(u)|_{_{u=u_0}}=0
\label{cbc}
\qqq for $u_0=r_0^{2-\xi\over 2}$. For non-integer $b$, the
corresponding eigen-functions of $N_2$ are 
\qq 
\varphi_{E,u_0}(u)\ =\ 
C_1(u_0)\, u^\hf\m J_{-b}(\tilde u)\,+\, C_2(u_0)\, 
u^\hf\m J_{b}(\tilde u)
\qqq
with $\tilde u\equiv{\sqrt{E/Z'}}\m u\m,$
\m see Eq.\,\,(\ref{egf}). Since \qq J_{{b}}(z)\ =\ 
{_{1}\over^{2^b\,\Gamma(1+{b})}}\,z^b\,(1+\CO(z^2))\m,
\label{asex}
\qqq the boundary condition (\ref{cbc}) implies that \qq
C_1(u_0)\,\CO(\tilde u_0) \,+\, C_2(u_0)\,\CO({\tilde u_0}^{2b-1})\ =\ 
0\m.  \qqq As a result, for (non-integer) $b<1$, $\,\lim\limits_{u_0\to0}
\,{_{C_2(u_0)}\over^{C_1(u_0)}}=\,0\,$ so that in the limit we obtain 
the eigen-functions of the operator $N_2^-$.
For (non-integer) $b>1$, however, $\,\lim\limits_{u_0\to0}\, 
{_{C_1(u_0)}\over^{C_2(u_0)}}=\,0\,$ and the eigen-functions tend
to those of $N_2^+$. The extension to the case of integer $b$ 
is equally easy.
\vskip 0.3cm

\nappendix{B}
%\label{A:two}
\vskip 0.5cm

Let us give here the explicit form of the integrals
of the kernels $\ee^{-t\m M^\mp_2}(r,r')$ against 
powers of $r'$. A direct calculation shows that for $\mu\geq 0$
and, respectively, $b<1$ and $b>-1$,
\qq
\int\limits_0^\infty\ee^{-t\m M_2^-}(r,r')
\,\,{r'}^{\mu}\,d\mu_d({r'})&=&
{_{\Gamma(1+{\mu\over2-\xi}-b)}
\over^{\Gamma(1-b)}}\,\left(4\m Z'\m t\right)^{\mu\over2-\xi}\cr
&&\cdot\,\,{}_{_1}F_{_1}(-{_\mu\over^{2-\xi}},
\m 1-b,\m-{_{r^{2-\xi}}\over^{4\m Z'\m
t}})\m,
\label{norm-}\\
\int\limits_0^\infty\ee^{-t\m M_2^+}(r,r')
\,\,{r'}^{\mu}\,d\mu_d({r'})&=&
{_{\Gamma(1+{\mu\over2-\xi})}
\over^{\Gamma(1+b)}}\,\left(4\m Z'\m
t\right)^{{\mu\over2-\xi}-b}\,\m r^{1-a}\cr
&&\cdot\,\,_{_1}F_{_1}(-{_{\mu}
\over^{2-\xi}}+b,\m 1+b,\m-{_{r^{2-\xi}}
\over^{4\m Z'\m t}})\m.
\label{norm+}
\qqq
A direct calculation gives also the asymptotic 
expansion of the kernels $\ee^{-t\m M_2^\mp}(r,r')$ 
at small $r$:
\qq
&&\ee^{-t\m M_2^{\mp}}(r,r')\ \ \ =\ \ \  
{_{2-\xi}\over^{S_{d-1}\m\Gamma(1\mp b)}}\,\left\{\matrix{
{r'}^{-d+2-\xi}\cr r^{1-a}\,{r'}^{-d+1+a-\xi}}\right\}\cr
&&\hspace{2.5cm}\cdot\ \sum\limits_{j=0}^\infty {_{(-1)^j\, 
r^{(2-\xi)j}}
\over^{j!\,(1\mp b)\cdots(j\mp b)\,(4\m Z'\m 
t)^{1+j}}}\ \,{_d\over^{(dz)^j}}\,(\m z^{j\mp b}\m
\ee^{-z}\m)\bigg\vert_{\m z\,=\m{{r'}^{2-\xi}
\over 4\m Z'\m t}}\m.\hspace{1.5cm}
\label{A12}
\qqq
An expansion around $r'=0$ may be obtained similarly.
\vskip 0.3cm

%\nappendix{3}
%\label{A:three}
%\vskip 0.5cm

%Another way to see that the exponential of the heat kernel $M_2^+$ is
%not normalized with respect to $r'$ (see eq. (\ref{nor+})) is to observe
%that the time derivative of $\ee^{-t\m M_2^\mp}(r,r')$ brings down the
%adjoint of $M_2^{\mp}$ acting on the $r'$ variable so that
%\qq
%{_{d}\over^{dt}}\,\m 
%\int\limits_0^\infty\ee^{-t\m M_2^\mp}(r,r')
%\,\,{r'}^{d-1}\m dr'&=&
%Z\int\limits_0^\infty\da_{r'}\m{r'}^a\da_{r'}
%\m{r'}^{\xi-a+d-1}\left(\ee^{-t\m M_2^\mp}(r,r')\right)\cr
%&=&Z\,{r'}^a\da_{r'}
%\m{r'}^{\xi-a+d-1}\left(\ee^{-t\m M_2^\mp}(r,r')\right)
%\bigg\vert^{_{r'=\infty}}_{^{r'=0}}\m.
%\qqq
%The contribution from $r'=\infty$ vanishes but
%\qq
%\ee^{-t\m M_2^\mp}(r,r')\ \sim\ \left\{\matrix{{r'}^{1+a-\xi-d}
%\cr{r'}^{2-\xi-d}}\right\}\cdot\m(1+\CO({r'}^{2-\xi}))
%\qqq
%so that the contribution from $r'=0$ is zero for $M_2^-$
%if $1+a-\xi>0$, which is the same condition 
%as $b<1$ or $\wp<{d\over\xi^2}$, but it is finite 
%for $M_2^+$.
%\vskip 0.3cm

\nappendix{C}
%\label{A:four}
\vskip 0.5cm

Here we shall consider the long time behavior of the integral
of the heat kernel of the operator $M_2^-$:
\qq
X(t,r,r')\ \equiv\ \smallint\limits_0^t 
\ee^{-s\m M_2^-}(r,r')\, ds\m.
\qqq
Using the explicit form (\ref{expo}) of the heat kernel
$\ee^{-s\m M_2^-}(r,r')$, we may rewrite the last definition as
\qq
X(t,r,r')\ =\ \smallint\limits_0^tds\smallint\limits_0^\infty
\ee^{-sE}\, E^{-{b}}\,G(E;r,r')\, dE\m,
\label{a21}
\qqq
where
\qq
G(E,r,r')\m=\m 
{_{1}\over^{(2-\xi)\m Z\m S_{d-1}}}\,r^{1-a\over 2}\,
E^{{b}}\, J_{-{b}}({\sqrt{E/Z'}}\m{r^{2-\xi\over2}})
\,\m J_{-{b}}({\sqrt{E/Z'}}\m{{r'}^{2-\xi\over2}})
\,\m {r'}^{-d+{3\over2}+{a\over 2}-\xi}\m.\ \quad
\qqq
Note that, by virtue of the relation (\ref{asex}),
$\,G(0,r,r')={_{(2-\xi)\m(4\m Z')^{{b}-1}}
\over^{\Gamma(1-{b})^2
\m S_{d-1}}}\,{r'}^{-d+1+a-\xi}\equiv G_0(r')\,$
and is independent of $r$. For finite times, 
the integration by parts, accompanier by the changes
of variable $sE\leftrightarrow E$ gives the following identity:
\qq
{b}\m X(t,r,r')\m=\m t^{b}\smallint\limits_0^\infty
\ee^{-E}\m E^{-{b}}\m G(t^{-1}E,r,r')\m dE\m+\m 
\smallint\limits_0^tds\smallint\limits_0^\infty
\ee^{-sE}\m E^{1-{b}}\m \da_{_E}G(E,r,r')\m dE\m.\quad
\label{intbp}
\qqq
For $b<0$, the $t\to\infty$ limit of $X(t,r,r')$
exists and defines the kernel $(M_2^-)^{-1}(r,r')$: 
\qq
\lim\limits_{t\to\infty}\,\,{b}\, X(t,r,r')&=&{b}
\smallint\limits_0^\infty E^{-{b}-1}\, G(E,r,r')\, dE\cr
&=&-\m{_1\over^{(2-\xi)\m Z\,S_{d-1}}}\cases{
\hbox to 4cm{${r'}^{-d+2-\xi}$\hfill}{\rm for\ }
r\leq r'\m,\cr
\hbox to 4cm{$r^{(2-\xi)b}\m{r'}^{-d+(2-\xi)(1-b)}$\hfill}
{\rm for\ }r\geq r'\m,}
\label{a24}
\qqq
compare to Eq.\,\,(\ref{35}). The last expression has been obtained 
by the direct integration and the condition $b<0$ was required 
by the convergence at zero of the $E$-integral (the kernels 
$(M_2^\pm)^{-1}(r,r')$ may be also found easily by gluing 
the zero modes of $M_2^\pm$). Note that the right 
hand side is a real analytic function of ${b}=
{{1-a}\over{2-\xi}}$. Now Eq.\,\,(\ref{intbp})
implies that
\qq
\lim\limits_{t\to\infty}\,\left[\m {b}\,X(t,r,r')\,-\,
\Gamma(1-{b})\, G_0(r')\, t^{b}\m\right]
\ =\ \smallint\limits_0^\infty
E^{-{b}}\, \da_{_E}G(E,r,r')\, dE
\qqq
exists for ${b}<1$ and is also real analytic in ${b}$.
On the other hand, it coincides with the long time limit in Eq.\,\,(\ref{a24})
for ${b}<0$ and, consequently, must be given by the right hand side
of this equation for all ${b}<1$. The same arguments work 
after the integration of the above expressions against 
$\chi(r')\m d\mu_d(r')$ and hence the convergence of the expression
(\ref{toA}) when $t\to\infty$.
\vskip 0.3cm

\nappendix{D}
%\label{A:five}
\vskip 0.5cm

It is easy to give the exact expressions for the stationary 
2-point functions of the scalars $\theta$ and $\rho$
for $\wp<{_{d-2+\xi}\over^{2\m\xi}}$ in the presence
of positive diffusivity $\kappa$. They are  
\qq
F^\theta_2(r)&=&{_1\over^{Z\, S_{d-1}}}\smallint\limits_r^\infty
f(r')\, g(r')\,\m\chi(r')\,\m d\mu_d(r')
\,+\,{_1\over^{Z\, S_{d-1}}}\,\m g(r)\smallint\limits_0^r
f(r')\,\m\chi(r')\,\m d\mu_d(r')\m,\label{2pkat}\\
F^\rho_2(r)&=&{_1\over^{Z\, S_{d-1}}}\,\m f(r)
\smallint\limits_r^\infty g(r')\,\m\chi(r')\,\m d\mu_d(r')
\,+\,{_1\over^{Z\, S_{d-1}}}\,\m f(r)\, g(r)\smallint\limits_0^r
\chi(r')\,\m d\mu_d(r')\m,
\label{2pkar}
\qqq
where
\qq
f(r)\m=\m(r^\xi+{_{2\m\kappa}\over^Z})^{^{-d+1+a-\xi\over\xi}}\qquad
{\rm and}\qquad g(r)\m=\m\smallint\limits_r^\infty f(\zeta)^{-1}
\,{_{\zeta^{-d+1}}\over^{\zeta^\xi+{2\m\kappa\over Z}}}\, d\zeta\m.
\qqq
In the limit $\kappa\to0$ these expressions pass into 
Eq.\,\,(\ref{35}) and (\ref{36}), respectively.
For ${_{d-2+\xi}\over^{2\m\xi}}<\wp<
{_d\over^{\xi^2}}$, similarly as at $\kappa=0$,
the 2-point functions $F_2^\theta(t,r)$ and $F_2^\rho(t,r)$
are pumped into the constant and the $f(r)$ zero modes of $M_2^\kappa$
and $(M_2^\kappa)^*$, respectively, and do not reach stationary limits, 
although the 2-point structure function of the tracer does.
For $\wp\geq{d\over\xi^2}$ the pumping into
the zero modes reaches a constant rate. In the limit 
$\kappa\to0$, the zero mode $f(r)$ into which $F_2^\rho$
is pumped goes to $r^{-d+1+a-\xi}$ for 
$\wp<{d\over\xi^2}$ but becomes the delta-function
$\delta(\Nr)$ for $\wp>{d\over\xi^2}$.
%\hfill
\eject
%\vskip 0.3cm

\nappendix{E}
%\label{A:six}
\vskip 0.5cm

Let us prove the explicit expression (\ref{rrsn}) for 
the even structure functions in the strongly compressible
regime. Note that
\qq
S^\theta_{_N}(t,r)\ =\ \sum\limits_{Q\subset\{1,\dots,N\}}
(-1)^{\vert Q^c\vert}\,\, F^\theta_{_N}(t,(\Nr)_{_{q\in Q}},
({\bf 0})_{_{q\in Q^c}})
\label{efsn}
\qqq
where $Q^c$ stands for the complement of $Q$.
In the strongly compressible phase, 
by the multiple application of Eq.\,\,(\ref{contr}),
we infer that
\qq
P^{t,s}_{_N}((\Nr)_{_{q\in Q}},({\bf 0})_{_{q\in Q^c}};\m
\un{\Nr'})&=&\int P^{t,s}_2(\Nr,{\bf 0};\m\Nr''_1,\Nr''_2)\cr
&&\cdot\ \prod\limits_{q\in Q}\delta(\Nr''_1-\Nr'_q)\,
\prod\limits_{q\in Q^c}\delta(\Nr''_2-\Nr'_q)\,\,d\Nr''_1\,\m
d\Nr''_2\m.
\label{mcontr}
\qqq
Substituting Eq.\,\,(\ref{Nc1}) into the expression
(\ref{efsn}) and using the relations (\ref{mcontr}), we
obtain
\qq
S^\theta_{_N}(t,r)&=&\sum\limits_{1\leq n<m\leq N}
\int\limits_0^tds\int P_2^{t,s}(\Nr,{\bf 0};\m\Nr''_1,\Nr''_2)\cr
&&\cdot\,\bigg(\sum\limits_{\{n,m\}\subset Q\subset\{1,\dots,N\}}
(-1)^{\vert Q^c\vert}\ F^\theta_{_{N-2}}(s,
(\Nr''_1)_{_{q\in Q\setminus\{n,m\}}},
(\Nr''_2)_{_{q\in Q^c}})\ \chi(0)\cr
&&\ \ +\sum\limits_{n\in Q\subset\{1,\dots,N\}\setminus\{m\}}
(-1)^{\vert Q^c\vert}\ F^\theta_{_{N-2}}(s,
(\Nr''_1)_{_{q\in Q\setminus\{n\}}},
(\Nr''_2)_{_{q\in Q^c\setminus\{m\}}})
\ \chi({\vert\Nr''_1-\Nr''_2\vert})\cr
&&\ \ +\sum\limits_{m\in Q\subset\{1,\dots,N\}\setminus\{n\}}
(-1)^{\vert Q^c\vert}\ F^\theta_{_{N-2}}(s,
(\Nr''_1)_{_{q\in Q\setminus\{m\}}},
(\Nr''_2)_{_{q\in Q^c\setminus\{n\}}})
\ \chi({\vert\Nr''_1-\Nr''_2\vert})\cr
&&\ \ +\sum\limits_{Q\subset\{1,\dots,N\}\setminus\{n,m\}}
(-1)^{\vert Q^c\vert}\ F^\theta_{_{N-2}}(s,(\Nr''_1)_{_{q\in Q}},
(\Nr''_2)_{_{q\in Q^c\setminus\{n,m\}}})\ \chi(0)\bigg)\,d\Nr''_1\,
d\Nr''_2\cr
\cr\cr
&=&\ N(N-1)\, \smallint\limits_0^tds\smallint\limits_0^\infty 
P_2^{t,s}(r,r')\,\,S^\theta_{_{N-2}}(s,r')\,\,(\chi(0)-\chi({r'}))\,\, 
d\mu_d({r'})\m,
\qqq
which is the sought relation.

\end{document}